\documentclass[preprint,aip,graphicx,cha,showpacs]{revtex4-1}

\usepackage{amsmath,amsfonts,bm,bbm}

\usepackage{float}
\usepackage{epsfig}
\usepackage{esint}
\usepackage{color}

\usepackage{multirow}

  \usepackage{paralist}
  \usepackage{epstopdf}
  \usepackage{graphics} 
 \usepackage[colorlinks=true]{hyperref}
 \hypersetup{urlcolor=blue, citecolor=red}
 \usepackage[latin1]{inputenc}
\usepackage[caption=false]{subfig}

 \usepackage{tikz-cd}

\usepackage{float}
\usepackage{epsfig}
\usepackage{esint}

\usepackage{bm,braket}

\newcommand{\pcb}[1]{\textcolor{black}{#1}}

 \newcommand{\U}{{\mathbf U}}

\newcommand{\R}{{\mathbb R}}

\renewcommand{\L}{{\mathbb L}}

\newcommand{\e}{{\rm e}}

\newcommand{\calP}{{\mathcal P}}
\newcommand{\calU}{{\mathcal U}}
\newcommand{\calV}{{\mathcal V}}
\newcommand{\calR}{{\mathcal R}}

\newcommand{\x}{\mathbf{x}}
\newcommand{\y}{\mathbf{y}}

\newcommand{\X}{\mathbf{X}}
\renewcommand{\P}{\mathbb{P}}

\renewcommand{\theequation}{\arabic{section}.\arabic{equation}}

\renewcommand{\u}{\mathbf u}
\newcommand{\E}{{\mathbb E}}
\newcommand{\Z}{\mathbb Z}

\newcommand{\calZ}{\mathcal Z}

\begin{document}
\title[Interacting particle systems with stochastic resetting]{Global density equations for interacting particle systems with stochastic resetting: from overdamped Brownian motion to phase synchronization}

\author{Paul C. Bressloff}
\address{Department of Mathematics, Imperial College London, London SW7 2AZ, UK.}

\date{\today}

\begin{abstract}

A wide range of phenomena in the natural and social sciences involve large systems of interacting particles, including plasmas, collections of galaxies, coupled oscillators, cell aggregations, and economic ``agents'. Kinetic methods for reducing the complexity of such systems typically involve the derivation of nonlinear partial differential equations for the corresponding global densities. In recent years there has been considerable interest in the mean field limit of interacting particle systems with long range interactions. Two major examples are interacting Brownian particles in the overdamped regime and the Kuramoto model of coupled phase oscillators. In this paper we analyze these systems in the presence of local or global stochastic resetting, where the position or phase of each particle independently or simultaneously resets to its original value at a random sequence of times generated by a Poisson process. In each case we derive the Dean-Kawasaki (DK) equation describing hydrodynamic fluctuations of the global density, and then use a mean field ansatz to obtain the corresponding nonlinear McKean-Vlasov (MV) equation in the thermodynamic limit. In particular, we show how the MV equation for global resetting is driven by a Poisson shot noise process, reflecting the fact that resetting is common to all of the particles and thus induces correlations that cannot be eliminated by taking a mean field limit. We then investigate the effects of local and global resetting on nonequilibrium stationary solutions of the macroscopic dynamics and, in the case of the Kuramoto model, the reduced dynamics on the Ott-Antonsen manifold. 
\end{abstract}
\maketitle

\noindent {\bf Large systems of interacting particles arise in a wide range of applications in the natural and social sciences. For example, in physics the particles could represent electrons or ions in a plasma, molecules in passive or active fluids, or galaxies in a cosmological model. On the other hand, particles in biological applications tend to be micro-organisms such as cells or bacteria that can exhibit non-trivial aggregation phenomena such as motility-based phase separation. Finally, in economics or social sciences, particles typically represent individual ``agents''. A major challenge is how to reduce the complexity of such systems. A classical approach is to derive a macroscopic model that provides a continuous description of the dynamics in terms of
global densities evolving according to non-linear partial differential equations.
Such kinetic formulations date back to the foundations of statistical mechanics and the Boltzmann equation 
of dilute gases interacting via direct collisions. In recent years, however, much of the focus has been on the mean field limit of particles with long range or collisionless interactions. In this paper we consider two paradigmatic examples, namely, interacting Brownian particles in the overdamped regime and the Kuramoto model of coupled phase oscillators. We develop the first systematic study of these systems in the presence of local or global stochastic resetting, where the position or phase of each particle independently or simultaneously resets to its original value at a random sequence of times generated by a Poisson process. In each case we derive Dean-Kawasaki (DK) equations describing hydrodynamic fluctuations of the global densities.  The DK equation plays an important role in the stochastic and numerical analysis of finite interacting particle systems, and provides a general framework for deriving mean field equations. Taking expectations of the DK equation with respect to the noise processes and applying a mean field ansatz, we obtain corresponding nonlinear McKean-Vlasov (MV) equations, which are used to investigate the effects of resetting on stationary solutions of the macroscopic dynamics.}


\section{Introduction}

The classical Dean-Kawasaki (DK) equation is a stochastic partial differential equation (SPDE) that describes fluctuations in the global density $\rho(\x,t)=N^{-1}\sum_{j=1}^N\delta(\x-\X_j(t)) $ of $N$ over-damped Brownian particles with positions $\X_j(t)\in \R^d$ at time $t$ \cite{Dean96,Kawasaki98}. Within the context of non-equilibrium statistical physics, the DK equation is commonly combined with dynamical density functional theory (DDFT) in order to derive hydrodynamical models of interacting particle systems \cite{Marconi99,Evans04,Archer04,Witt21}. 
It is an exact equation for the global density (or empirical measure) in the distributional sense, and plays an important role in the stochastic and numerical analysis of interacting particle systems \cite{Dirr16,Konarovskyi19,Konarovskyi20,Djurdjevac22a,Djurdjevac22b,Fehrman23,Cornalba23}. There is also considerable mathematical interest in the rigorous stochastic analysis of the mean field limit $N\rightarrow \infty$ for overdamped Brownian particles with weak interactions, see for example Refs. \onlinecite{Oelsch84,Jabin17,Pavliotis21,Chaintron22a,Chaintron22b}. 
In particular, if the initial positions of the $N$ particles are independent and identically distributed, then for a wide range of systems it can be proven that $ \rho $ converges in distribution to the solution of the McKean-Vlasov (MV) equation \cite{McKean66}; the latter is a nonlocal nonlinear Fokker-Planck (FP) equation for the mean field density.
The interacting particle system is said to satisfy the propagation of chaos property. The MV equation can also be derived directly from the DK equation by taking expectations with respect to the independent white noise processes and imposing a mean field ansatz. The MV equation is of interest in its own right, since it can support multiple stationary solutions and associated phase transitions \cite{Tamura84,Tugaut14}. This has been explored in various configurations, including double-well confinement and Curie-Weiss (quadratic) pairwise interactions on $\R$ \cite{Desai78,Dawson83,Pavliotis19}, and interacting particles on a torus \cite{Chayes10,Carrillo20}. 

A well-known example of the latter is the stochastic Kuramoto model of interacting phase oscillators with sinusoidal coupling and quenched disorder due to the random distribution of natural frequencies \cite{Kuramoto84,Strogatz00,Acebron05}. The corresponding global density is $\rho(\theta,t,\omega)=N^{-1}\sum_{j=1}^N\delta (\theta-\Theta_j(t))\delta(\omega-\omega_j)$ where $\Theta_j(t)$ and $\omega_j$ are the stochastic phase and natural frequency of the $j$th oscillator. The well-known continuum model for the density of phase oscillators \cite{Sakaguchi88,Strogatz91,Crawford94,Crawford99} is precisely the MV equation for the global density in the mean field limit $N\rightarrow \infty$, whose existence can be proven rigorously using propagation of chaos \cite{Dai96}.

In this paper we study interacting Brownian particles and Kuramoto phase oscillators in the presence of stochastic resetting. That is, the position $\X_j(t)$ or phase $\Theta_j(t)$ of the $j$th particle resets to its initial value $\x_{j,0}$ or $\theta_{j,0}$ at a random sequence of times generated from a Poisson process with constant rate $r$. We consider two distinct types of resetting protocol. (A) All of the particles simultaneously reset (global resetting). (B) Each particle independently resets according to its own sequence
of resetting times (local resetting). In each case we derive a generalized DK equation for the global density, and then use a mean field ansatz to derive a corresponding MV equation. The MV equation for local resetting is obtained by averaging the DK equation with respect to the white noise and resetting processes, and is thus deterministic. On the other hand, the MV equation for global resetting depends stochastically on the global resetting times, since we only average the DK equation with respect to the white noise processes. The latter reflects the fact that resetting is common to all of the particles and thus induces correlations that cannot be eliminated by taking a mean field limit. An analogous result applies to systems of particles subject to a randomly switching environment \cite{Bressloff24}.

Although most previous studies of stochastic resetting have focused on single-particle systems \cite{Evans11a,Evans11b,Evans14,Evans20}, there is a growing interest in understanding the effects of global or local resetting on interacting particle systems (see the recent review of Ref. \onlinecite{Nagar23}). Resetting introduces an additional timescale into the dynamics and an ordering mechanism (assuming the particles reset to neighboring locations) that competes with disorder due to diffusion and possible cooperative effects due to particle interactions. As far as we are aware, current work on multi-particle systems with resetting has concentrated on either small numbers of particles \cite{Falcao17,Evans22} eg. predator-prey models, or large lattice systems such as symmetric exclusion processes \cite{Basu19,Sadekar20,Miron21}, totally asymmetric exclusion processes \cite{Karthika20,Pelizzola21}, and aggregation processes \cite{Durang14,Grange21}. There has been one recent study of the Kuramoto model with global resetting \cite{Sarkar22}, which will be incorporated into the more general theoretical framework developed here.

The structure of the paper is as follows. In Sec. II we begin by briefly reviewing the theory of interacting overdamped Brownian particles without resetting. We then use It\^o's lemma to derive the generalized DK equation in the presence of global or local resetting, and show how to obtain the corresponding MV equation by taking averages and using a mean field ansatz. In Sec. III we illustrate the effects of local resetting by analyzing stationary solutions of the one-dimensional (1D) deterministic MV equation with Curie-Weiss (quadratic) pairwise interactions. The analysis is complicated by the fact that the stationary solution in the absence of interactions is a nonequilibrium stationary state (NESS) rather than a Boltzmann distribution. In the presence of interactions, the NESS depends self-consistently on its own first moment. In Sec. IV we show how global resetting induces statistical correlations even in the absence of particle interactions. The global density for noninteracting Brownian particles satisfies a stochastic linear FP equation driven by a Poisson shot noise process due to resetting. We derive moment equations for the associated probability density functional, and thus establish the existence of two-point correlations. 

In Sec. V, we develop the analogous theory for the Kuramoto model with either local or global resetting. We first derive the  DK equation with quenched disorder in the natural frequencies, and then take averages with respect to the noise processes using a mean field ansatz to obtain the corresponding MV equations. In the case of  local resetting, we show in an appendix how there is a break down of the Ott-Antonsen (OA) theory for the classical Kuramoto model \cite{Ott08}, that is, the dynamics cannot be restricted to the two-dimensional OA manifold. Nevertheless, interpolating between the limiting cases $r\rightarrow 0$ and $r\rightarrow \infty$, we establish how  local resetting smooths the classical phase transition from an incoherent to a partially coherent state, which is now represented by an`` imperfect bifurcation''. Finally, applying the OA theory to the stochastic MV equation for global resetting leads to a piecewise deterministic dynamical system with reset on the low-dimensional OA manifold. We thus recover the stochastic OA dynamical model analyzed in Ref. \onlinecite{Sarkar22}. Stationary solutions are now specified in terms of the NESS of the reset dynamics on the OA manifold.

\vfill

\section{Interacting Brownian particles}

A classical result in statistical physics is that for a finite system of overdamped Brownian particles subject to conservative inter-particle and external forces, the corresponding linear Fokker-Planck (FP) equation has a unique stationary solution given by the Boltzmann distribution. More specifically, let $\X_j(t)\in \R^d$ denote the position of the $j$th particle at time $t$, $j=1,\ldots,N$. Suppose that the positions evolve according to the stochastic differential equation (SDE)
\begin{equation}
\label{SDE0}
d\X_j(t)=-\gamma^{-1}{\bm \nabla}_j U(\X_1(t),\ldots ,\X_N(t))dt+\sqrt{2D}d{\bf W}_j(t),
\end{equation}
where $U$ is the multi-particle potential, $D$ is the diffusivity, $\gamma$ is a drag coefficient satisfying the Einstein relation $D\gamma =k_BT$, and ${\bf W}_j(t)$ is a vector of independent Wiener processes. In addition, the subscript on ${\bm \nabla}_j$ indicates that differentiation is with respect to $\X_j$. The corresponding multivariate FP equation for the joint probability density $p(\x_1,\ldots,\x_N,t)$ is
\begin{equation}
\frac{\partial p}{\partial t}=D\sum_{j=1}^N{\bm \nabla}^2_j p+\frac{1}{\gamma}\sum_{j=1}^N{\bm \nabla}_j\cdot \left ({\bm \nabla}_j U(\x_1,\ldots ,\x_N)p\right ),
\end{equation}
which has the unique stationary solution $p(\x_1,\ldots\x_N)=Z^{-1}\exp(-\beta U(\x_1,\ldots,\x_N))$ with $\beta =1/(k_BT)$. In the case of noninteracting particles subject to a common external potential $V$, the multi-particle potential is \begin{equation}
 U_1(\x_1,\ldots,\x_N)\equiv \sum_{j=1}^NV(\x_j),
 \end{equation}
and the stationary solution takes the product form
  \begin{equation}
  p(\x_1,\ldots,\x_N)=\prod_{j=1}^N \rho(\x_j),\quad \rho(\x)=Z_{\rm 1p}^{-1}\e^{-\beta V(\x)},
  \end{equation}
   where $Z_{\rm 1p}$ the one-particle partition function. The most common type of interaction potential is a pairwise potential:
\begin{equation}
U_2(\x_1,\ldots,\x_N)\equiv \frac{1}{2N}\sum_{j,k=1}^N K(\x_j-\x_k),\quad K(0)=0,
\end{equation}
with ${\bm \nabla}_jU_2(\x_1,\ldots,\x_N)= N^{-1}\sum_{k=1}^N{\bm \nabla}K(\x_j-\x_k)$.

The existence of a unique stationary density for the finite system reflects the fact that the dynamics is ergodic. However, ergodicity may break down in the thermodynamic limit $N\rightarrow \infty$, resulting in the coexistence of multiple stationary states and their associated phase transitions. One way to explore this issue is to consider a ``hydrodynamic'' formulation of the multi-particle Brownian gas, which is based on the stochastic dynamics of the global density (or emprical measure)
\begin{equation}
\label{global}
\rho(\x,t)=N^{-1}\sum_{j=1}^N\delta(\x-\X_j(t)) .
\end{equation}
Using the standard form for the multi-particle potential, namely $U=U_1+U_2$, Eq. (\ref{SDE0}) becomes
 \begin{align}
 \label{SDE1}
  d\X_j(t)&=-\frac{1}{\gamma}\bigg [{\bm \nabla}V(\X_j(t))+\frac{1}{N}\sum_{k=1}^N{\bm \nabla} K(\X_j(t)-\X_k(t))\bigg ]dt 
  +\sqrt{2D}d{\bf W}_j(t).
 \end{align}
It can be shown that $\rho$ evolves according to the Dean-Kawasaki (DK) equation \cite{Dean96,Kawasaki98} (see also Sec. IIB)
 \begin{align}
  \frac{\partial \rho(\x,t)}{\partial t} 
&=\sqrt{\frac{2D}{N}}{\bm \nabla} \cdot \bigg [ \sqrt{ \rho(\x,t)} { \bm \eta}(\x,t)\bigg ]+D{\bm \nabla}^2 \rho(\x,t) \nonumber \\
 & \quad  +\frac{1}{\gamma} {\bm \nabla}\cdot \bigg (  \rho(\x,t) \int_{\R^d}    \rho(\y,t){\bm \nabla}K(\x-\y)d\y \bigg ) ,
\label{DKN}
\end{align}
where ${\bm \eta}(\x,t)$ is a vector of independent spatiotemporal white noise processes. 
Formally speaking, Eq. (\ref{DKN}) is an exact equation for the global density in the distributional sense. 

It is clear that averaging the DK Eq. (\ref{DKN}) with respect to the white noise processes results in a moment closure problem for the one-particle density $\langle \rho(\x,t)\rangle$, since it couples to the two-point correlation function $\langle \rho(\x,t)\rho(\y,t)\rangle$. The latter itself couples to the three point correlation functions etc. One way to achieve moment closure is to take the thermodynamic limit $N\rightarrow \infty$ and use mean field theory, see for example Refs.  \onlinecite{Oelsch84,Jabin17,Pavliotis21,Chaintron22a,Chaintron22b}. (This is possible because we have scaled the pairwise potential $U_2$ by a factor of $1/N$.) In particular, suppose that the initial positions of the $N$ particles are independent and identically distributed, i.e. the joint probability density at $t=0$ takes the product form
$p(\x_1,\ldots,\x_{N},0)=\prod_{j=1}^{N}\phi_0(\x_j)$. Under mild conditions on the potential functions $V$ and $K$, it can then be proven that $ \rho(\x,t) $ converges in distribution to a solution $\phi(\x,t)$ of the so-called McKean-Vlasov (MV) equation \cite{McKean66} in the limit $N\rightarrow \infty$:
\begin{align}
  \frac{\partial \phi(\x,t)}{\partial t} 
&=D{\bm \nabla}^2 \phi(\x,t)  +\frac{1}{\gamma}{\bm \nabla}\cdot  \bigg ( \phi(\x,t) \bigg [{\bm \nabla }V(\x) +\int_{\R^d} \phi(\y,t) {\bm \nabla}K(\x-\y)d\y\bigg ] \bigg ),
\label{MVr0}\end{align}
with $\phi(\x,0)=\phi_0(\x)$. The interacting Brownian gas is said to satisfy the propagation of chaos property. 
Eq. (\ref{MVr0}) takes the form of a nonlinear nonlocal FP equation for the so-called McKean SDE
\begin{align}
 d\X&=-\frac{1}{\gamma} \left [{\bm \nabla}V(\X(t))+\int_{\R^d}{\bm \nabla }K(\X(t)-\y)\phi(\y,t)d\y\right ]dt 
 +\sqrt{2D}d{\bf W}(t).
\end{align}
The MV equation is the starting point for exploring the existence of multiple stationary solutions and associated phase transitions for an infinite system of interacting Brownian particles \cite{Tamura84}. Examples include double-well confinement and Curie-Weiss interactions on $\R$ \cite{Desai78,Dawson83,Pavliotis19}, and interacting particles on a torus \cite{Chayes10,Carrillo20}. In the specific case of the Curie-Weiss potential $K(\x-\y)=\lambda (\x-\y)^2/2$, the coupling term in the SDE (\ref{SDE1}) becomes $-\lambda(\X_j(t)-\overline{\X}(t))$ where $\overline{\X}(t)=N^{-1}\sum_{k=1}^N\X_k(t)$.
It is an example of a cooperative coupling that tends to make
the system relax towards the ``center of gravity'' of the multi-particle ensemble. If $V(\x)$ is given by a multi-well potential then there is competition between the cooperative interactions and the tendency of particles to be distributed across the different potential wells according to the classical Boltzmann distribution.

\subsection{SDE with stochastic resetting}

We now incorporate stochastic resetting by assuming that each of the Brownian particles can reset to its initial position $\x_{0,j}$ at a fixed Poisson rate $r$. Following Ref. \onlinecite{Nagar23}, we distinguish between a global update rule where all of the particles simultaneously reset and a local resetting scheme in which the particles independently reset. Let $T_{j,n}$ denote the $n$th resetting time of the $j$th particle with $n\geq 0$. In the case of local resetting, the inter-reset intervals $\tau_{j,n}=T_{j,n+1}-T_{j,n}$ are identical independently distributed random variables for all $j,n$ with $\P[\tau_{j,n}\in [\tau,\tau+d\tau]]=r\e^{-r\tau}d\tau$. On the other hand, in the case of global resetting we have $T_{j,n}=T_n$ for all $j=1,\ldots,N$ with $\tau_n=T_{n+1}-T_n$ exponentially distributed. 

Given the sequence of resetting times, the SDE (\ref{SDE0}) becomes
\begin{align}
 \frac{d\X_j(t)}{dt}&=\gamma^{-1}{\bm \nabla}_j U(\X_1(t),\ldots ,\X_N(t))+\sqrt{2D}{\bm \xi}_j(t) +\sum_{n=1}^{\infty}(\x_{j,0}-\X_j(t))\delta(t-T_{j,n}).
 \label{SDEres}
\end{align}
We have formally set $d{\bf W}_j(t)={\bm \xi}_j(t)dt$ where $ {\bm \xi}_j$ is a $d$-dimensional white noise term such that
\begin{equation}
\langle {\bm \xi}_i(t)\rangle =0,\quad \langle {\xi}_i^{\sigma}(t){\xi}_j^{\sigma'}(t')\rangle =\delta(t-t')\delta_{i,j}\delta_{\sigma,\sigma'}.
\end{equation}
Integrating Eq. (\ref{SDEres}) over an interval $[t,t+dt]$ implies that
\begin{subequations}
\label{SDEres2}
\begin{align}
d\X_j(t)&=\gamma^{-1}{\bm \nabla}_j U(\X_1(t),\ldots ,\X_N(t))dt +\sqrt{2D}d{\bf W}_j(t), \quad T_{j,n}\notin [t,t+dt],\\
d\X_j(t)&=x_{j,0}-\X_j(t),\quad T_{j,n} \in [t,t+dt].
\end{align}
\end{subequations}
with $d\X_j(t)=\X_j(t+dt)-\X_j(t)$. 
In standard formulations of stochastic resetting at the single-particle level \cite{Evans20}, the SDE (\ref{SDEres2}) is averaged over multiple realizations of the Poisson resetting process. Since $\P[T_{j,n}\in [t,t+dt]]=rdt$, Eq. (\ref{SDEres2}) becomes
\begin{subequations}
\label{SDEres3}
\begin{align}
d\X_j(t)&=\gamma^{-1}{\bm \nabla}_j U(\X_1(t),\ldots ,\X_N(t))dt   +\sqrt{2D}d{\bf W}_j(t) \mbox{ with probability } 1-rdt,\nonumber\\
d\X_j(t)&=x_{j,0}-\X_j(t) \mbox{ with probability } rdt.\end{align}
\end{subequations}
However, at the multi-particle level, in order to distinguish between local and global resetting it is necessary to work with the doubly stochastic SDE (\ref{SDEres}), whereby resetting is modeled as a Poisson shot process.

\subsection{Dean-Kawasaki equation} We begin by deriving the DK equation for the global density (\ref{global}) in the presence of resetting by generalizing the construction of Ref. \onlinecite{Dean96}. Consider an arbitrary smooth test function $f: \R^d\rightarrow \R$. \pcb{Set $N^{-1}\sum_{j=1}^Nf(\X_j(t))=\int_{\R^d} \rho(\x,t)f(\x)d\x$ so that
\begin{align*}
& \int_{\R^d}d\x\,  f(\x)\frac{\partial \rho(\x,t)}{\partial t}    =\lim_{\Delta t\rightarrow 0}\frac{1}{N}\sum_{j=1}^N\frac{f(\X_j(t+\Delta t))-f(\X_j(t))}{\Delta t}    \end{align*}
 with}
\begin{align*}
  \X_j(t+\Delta t)&=\X_j( t)-\gamma^{-1}{\bm \nabla}_j U(\X_1(t),\ldots ,\X_N(t))\Delta t +\sqrt{2D}\Delta {\bf W}_j(t)
+\sum_{n=1}^{\infty}(\x_{j,0}-\X_j(t))\delta_{t,T_{j,n}}.
\end{align*}
Using It\^o's lemma to Taylor expand $f(\X_j(t+dt))$ about $\X_j(t)$, we have
 \begin{align*}\Delta f
  & ={\bm \nabla}f(\X_j(t))\Delta \X_j(t)+ {\bm \nabla}^2f(\X_j(t))\Delta \X_j(t)^2+\ldots \nonumber \\
 & = -\gamma^{-1}{\bm \nabla}f(\X_j(t)) {\bm \nabla}_jU(\X_1(t),\ldots ,\X_N(t))\Delta t  +\sqrt{2D}{\bm \nabla}f(\X_j(t)) \Delta {\bf W}_j+ D{\bm \nabla}^2f(\X_j(t))\Delta t  \nonumber \\
 &\quad +  \bigg [f(\x_{0,j})-f(\X_j(t))\bigg ]\sum_{n\geq 1}  \delta_{t,T_{j,n}}+o(\Delta t).
\end{align*}
\pcb{Taking the limit $\Delta t \rightarrow 0$, setting $U=U_1+U_2$ and using the definition of $\rho$ gives
\begin{align*}
 & \int_{\R^d}d\x\, f(\x)\frac{\partial \rho(\x,t)}{\partial t} 
   \nonumber \\&=\int_{\R^d}d\x\, \bigg [\frac{\sqrt{2D}{\bm \nabla} f(\x)}{N}\cdot\sum_{j=1}^N\rho_j(\x,t)  {\bm \xi}_j(t)+  \rho(\x,t)\bigg (D {\bm \nabla}^2 f(\x)
   -{\bm \nabla} f(\x)\cdot \calV[\x,t,\rho]\bigg )\nonumber \\
 &\quad +\frac{1}{N}\sum_{j=1}^Nh_j(t)  [\delta(\x-\x_{0,j})-\rho_j(\x,t)]f(\x)\bigg ],\nonumber
\end{align*}
where $\rho_j(\x,t)=\delta(\x-\X_j(t))$,
\begin{equation}
\label{calV}
\calV[\x,t,\rho]=\frac{1}{\gamma}\bigg [ {\bm \nabla}V(\x) + \int_{\R^d}d\y \,  \rho(\y,t){\bm \nabla} K(\x-\y)   \bigg ].
\end{equation}
 and}
\begin{equation}
\label{shot}
h_j(t)=\sum_{n\geq 1} \delta(t-T_{j,n})
\end{equation}
represents a Poisson shot noise process.
Integrating by parts the various terms involving derivatives of $f$ and using the fact that $f$ is arbitrary yields the following SPDE for $\rho$:
\begin{subequations}
\label{rho1}
\begin{align}
 \frac{\partial \rho(\x,t)}{\partial t} 
 &=-\sqrt{\frac{2D}{N^2}}\sum_{j=1}^{N}{\bm \nabla} \cdot \bigg [ \rho_j(\x,t) {\bm \xi}_j(t)\bigg ]+D{\bm \nabla}^2  \rho(\x,t)  + {\bm \nabla} \cdot \bigg (\rho(\x,t) \calV[\x,t,\rho]\bigg ) \\
 &\quad +\frac{1}{N}\sum_{j=1}^N \bigg [\delta(\x-\x_{0,j})-\rho_j(\x,t)\bigg ]h_j(t)\nonumber
\end{align}
\end{subequations}

As it stands, Eq. (\ref{rho1}) is not a closed equation for $\rho$ due to the noise terms. Following Ref. \onlinecite{Dean96},  we introduce the space-dependent Gaussian noise term
\begin{equation}
{ \xi}(\x,t)=-\frac{1}{N}\sum_{j=1}^{N}{\bm \nabla} \cdot \bigg [ \rho_j(\x,t) {\bm \xi}_j(t)\bigg ]
\end{equation}
with zero mean and the correlation function
\begin{equation*}
\langle 
\xi (\x,t)\xi (\y,t')\rangle = \frac{\delta(t-t')}{N^2}\sum_{j=1}^{N} {\bm \nabla}_{\x}\cdot {\bm \nabla}_{\y}\bigg (\rho_j(\x,t) \rho_j(\y,t) \bigg ).
\end{equation*}
Since $\rho_j(\x,t) \rho_j(\y,t) =\delta(\x-\y)\rho_j(\x,t)$, it follows that
\begin{equation*}
\langle 
\xi (\x,t)\xi (\y,t')\rangle = \frac{1}{N}\delta(t-t') {\bm \nabla}_{\x}\cdot {\bm \nabla}_{\y}\bigg (\delta(\x-\y)\rho(\x,t) \bigg ).
\end{equation*}
Finally, we introduce the global density-dependent noise field
\begin{equation}
\widehat{\xi }(\x,t)=\frac{1}{\sqrt{N}}{\bm \nabla} \cdot\bigg ({\bf \eta}(\x,t)\sqrt{\rho}(\x,t)\bigg ),
\end{equation}
where ${\bf \eta}(\x,t)$ is a global white noise field whose components satisfy 
\begin{equation}
\langle  \eta^{\sigma}(\x,t)\eta^{\sigma'}(\y,t')\rangle =\delta(t-t')\delta(\x-\y)\delta_{\sigma,\sigma'}.
\end{equation}
It can be checked that the Gaussian noises ${ \xi}$ and $\widehat{\xi}$ have the same correlation functions and are thus statistically identical. We thus obtain a modified version of the classical DK equation:
\begin{align}
 \frac{\partial \rho(\x,t)}{\partial t} 
&=\sqrt{\frac{2D}{N}}{\bm \nabla} \cdot \bigg [ \sqrt{\rho(\x,t)} {\bm \eta}(\x,t)\bigg ]+D{\bm \nabla}^2  \rho(\x,t)\ +{\bm \nabla} \cdot \bigg (\rho(\x,t) \calV[\x,t,\rho]\bigg )\label{rhoc} \\
&\quad +\frac{1}{N}\sum_{j=1}^N \bigg [\delta(\x-\x_{0,j})-\rho_j(\x,t)\bigg ]h_j(t).\nonumber \end{align}
Setting $h_j=0$ (no resetting) recovers the classical DK Eq. (\ref{DKN}). However, in the presence of local resetting the DK equation (\ref{rhoc}) is no longer a closed SPDE for the global density $\rho(\x,t)$ due to the shot noise terms. On the other hand, a closed SPDE is obtained in the case of global resetting, since $h_j(t)=h(t)=\sum_n\delta(t-T_n)$ for all $j$ and Eq. (\ref{rhoc}) becomes
\begin{align}
 \frac{\partial \rho(\x,t)}{\partial t} 
&=\sqrt{\frac{2D}{N}}{\bm \nabla} \cdot \bigg [ \sqrt{\rho(\x,t)} {\bm \eta}(\x,t)\bigg ]+D{\bm \nabla}^2  \rho(\x,t)  +{\bm \nabla} \cdot \bigg (\rho(\x,t) \calV[\x,t,\rho]\bigg ) \nonumber \\
&\quad +\pcb{\left [\frac{1}{N}\sum_{j=1}^N\delta(\x-\x_{j,0})-\rho(\x,t) \right ]h(t) .}\label{rhoglob}\end{align}

\subsection{McKean-Vlasov equations} Irrespective of global vs local resetting, Eq. (\ref{rhoc}) suffers from the same problem as the classical DK Eq. (\ref{DKN}), namely, averaging with respect to the noise processes results in a moment closure problem for the one-particle density $\langle \rho\rangle$. Therefore, in order to make further progress, we will impose a mean field ansatz. The details, however, will depend on whether resetting is local or global. 
\medskip

\noindent {\bf Local resetting.} In this scenario we take expectations of Eq. (\ref{rhoc}) with respect to the white noise and resetting processes. \pcb{Let
$\phi(\x,t)=\bigg \langle \E[\rho(\x,t)]\bigg \rangle$,
where $\E[\cdot]$ and $\langle \cdot \rangle$ denote averaging with respect to the resetting and white noise processes, respectively. We then have
\begin{align}
  \frac{\partial \phi(\x,t)}{\partial t} 
&=D{\bm \nabla}^2  \phi(\x,t)+\frac{1}{\gamma}{\bm \nabla} \cdot \bigg (\phi(\x,t) {\bm \nabla}V(\x)\bigg )  \\
&\quad +\frac{1}{\gamma}{\bm \nabla} \cdot \bigg (\int_{\R^d} {\bm \nabla}K(\x-\y)  \bigg \langle \E[\rho(\x,t)\rho(\y,t)]\bigg \rangle\bigg )-r\left [\phi(\x,t)-\frac{1}{N}\sum_{j=1}^N\delta(\x-\x_{j,0})\right ] .\nonumber
\end{align}
The last line follows from the independence of the resetting process on particle positions and Campbell's theorem for shot noise processes:
\begin{align}
 \E\bigg [\frac{1}{N}\sum_{j=1}^N \bigg (\delta(\x-\x_{0,j})-\rho_j(\x,t)\bigg )h_j(t)\bigg ] &=
\frac{1}{N}\sum_{j=1}^N \E\bigg [\delta(\x-\x_{0,j})-\rho_j(\x,t) \bigg ] \E[h_j(t) ]\nonumber \\
 &\quad =r\left [\frac{1}{N}\sum_{j=1}^N\delta(\x-\x_{j,0})-\rho(\x,t) \right ].
\end{align}
As it stands, we still have a moment closure problem, since $\phi(\x,t)$ couples to the two-point correlation function, which in turn couples to the three-point correlation function etc. Therefore, we now take the thermodynamic limit $N\rightarrow \infty$ under
the mean field ansatz
\begin{align}
 \bigg \langle \E[\rho(\x,t)\rho(\y,t)]\bigg \rangle&=\bigg \langle \E[\rho(\x,t)]\bigg \rangle\bigg \langle \E[\rho(\y,t)]\bigg \rangle
=\phi(\x,t)\phi(\y,t).
\end{align}
}
We thus obtain the deterministic MV equation
\begin{subequations}
\label{MVloc}
\begin{align}
  \frac{\partial \phi(\x,t)}{\partial t} 
&=D{\bm \nabla}^2  \phi(\x,t)+{\bm \nabla} \cdot \bigg (\phi(\x,t) \calV[\x,t,\phi]\bigg ) -r\phi(\x,t)+r\rho_0(\x) ,
\end{align}
with
\begin{equation}
\label{calVr}
\calV[\x,t,\phi]=\frac{1}{\gamma}\bigg [ {\bm \nabla}V(\x) + \int_{\R^d}d\y \,  \phi(\y,t){\bm \nabla} K(\x-\y)   \bigg ].
\end{equation}
\end{subequations}
and
\begin{equation}
\pcb{\rho_0(\x) =\lim_{N\rightarrow \infty} \frac{1}{N}\sum_{j=1}^N\delta(\x-\x_{j,0}).}
\end{equation}
In the absence of stochastic resetting, Eq. (\ref{MVloc}) reduces to the classical MV Eq. (\ref{MVr0}). In Sec. IIII we will consider a 1D version of Eq. (\ref{MVloc}) with Curie-Weiss interactions in order to analyze stationary solutions in the presence of local resetting.
\medskip

\noindent {\bf Global resetting.} In the case of global resetting, we have a closed DK equation for $\rho(\x,t)$ but we can no longer take expectations of Eq. (\ref{rhoglob}) with respect to the global shot noise process $h(t)$, since it is common to all of the particles. Averaging with respect to the white noise processes, setting
$\Phi(\x,t)=\langle \rho(\x,t)\rangle$, 
and imposing the mean field ansatz
\begin{equation}
 \bigg \langle \rho(\x,t)\rho(\y,t)\bigg \rangle= \langle  \rho(\x,t) \rangle  \langle  \rho(\y,t) \rangle
=\Phi(\x,t)\Phi(\y,t),
\end{equation}
now yields the stochastic MV equation 
\begin{subequations}
\label{MVglob}
\begin{align}
  \frac{\partial \Phi(\x,t)}{\partial t} 
&=D{\bm \nabla}^2  \Phi(\x,t)+{\bm \nabla} \cdot \bigg (\Phi(\x,t) \calV[\x,t,\Phi]\bigg ) +[\rho_0(\x)-\Phi(\x,t)]h(t),
\end{align}
with
\begin{equation}
\label{calVrr}
\calV[\x,t,\Phi]=\frac{1}{\gamma}\bigg [ {\bm \nabla}V(\x) + \int_{\R^d}d\y \,  \Phi(\y,t){\bm \nabla} K(\x-\y)   \bigg ].
\end{equation}
\end{subequations}
We see that the MV equation is driven by external Poisson shot noise $h(t)$. In Sec. IV we will show how this common drive results in statistical correlations even in the absence of particle interactions, analogous to the MV equation for Brownian particles driven by a randomly switching environment \cite{Bressloff24}.

As we highlighted at the beginning of Sec. II, in the absence of stochastic resetting, the validity of the mean field ansatz can be proven using propagation of chaos \cite{Oelsch84,Jabin17,Pavliotis21,Chaintron22a,Chaintron22b}. The latter is essentially a version of the law of large numbers, so that simulations for large but finite $N$ generate macroscopic quantities that are consistent with solutions to the deterministic MV equation up to $O(1/\sqrt{N})$ errors. In the case of local resetting, we assume that the mean field limit also holds when $r>0$. That is, we take the independent resetting of a large number of particles to be self-averaging so that Campbell's theorem can be applied to the individual shot noise processes. This is no longer possible in the case of global resetting. Nevertheless, we assume that the classical mean field limit applies in between resetting events, and that this is preserved by the initial conditions immediately after each reset. Since the resetting times are stochastic, it follows that the corresponding MV equation is driven by a Poisson shot noise process. In other words, the behavior of the infinite particle system is no longer deterministic. One final comment is that the global resetting rule $\X_j(t)\rightarrow \x_{j,0}$ for all $j=1,\ldots ,N$ implies that the initial condition immediately after reset is fixed, rather than randomly distributed according to some chaotic law. It is known that the mean field limit with deterministic initial data can break down in the case of singular interaction kernels $K$ such as the Coulomb potential, for which $|K(\x)|\rightarrow \infty$ as $\x\rightarrow 0$. For example, it is easy to construct initial data in which the particles are highly concentrated so that interaction forces are extremely strong. In this paper we only consider non-singular kernels that are Lipschitz-bounded.

\setcounter{equation}{0}
\section{Stationary solutions of the 1D McKean-Vlasov equation (local resetting)}

Establishing the existence of stationary solutions of the deterministic MV Eq. (\ref{MVloc}) is non-trivial, even in the absence of resetting. In this section we explore this issue in terms of the following 1D model with Curie-Weiss coupling:
\begin{align}
 \frac{\partial \phi(x,t)}{\partial t}
&=D\frac{\partial^2 \phi(x,t)}{\partial x^2}+ \frac{\partial}{\partial x}\bigg (\phi(x,t)\calV[x,t,\phi]\bigg )-r\phi(x,t)+r\delta(x-x_0) .
\label{MV1}
\end{align}
with
\begin{equation}
\calV[x,t,\phi]=\frac{1}{\gamma}\bigg [ V'(x)+\lambda \int_{-\infty}^{\infty}(x-y)  {\phi}(y,t) dy\bigg ].
\end{equation}
For simplicity, we also assume that $x_{0,j}=x_0$ for all $j=1,\ldots,N$ so that $\rho_0(x)=\delta(x-x_0)$.
\medskip

\begin{figure}[b!]
\centering
\includegraphics[width=10cm]{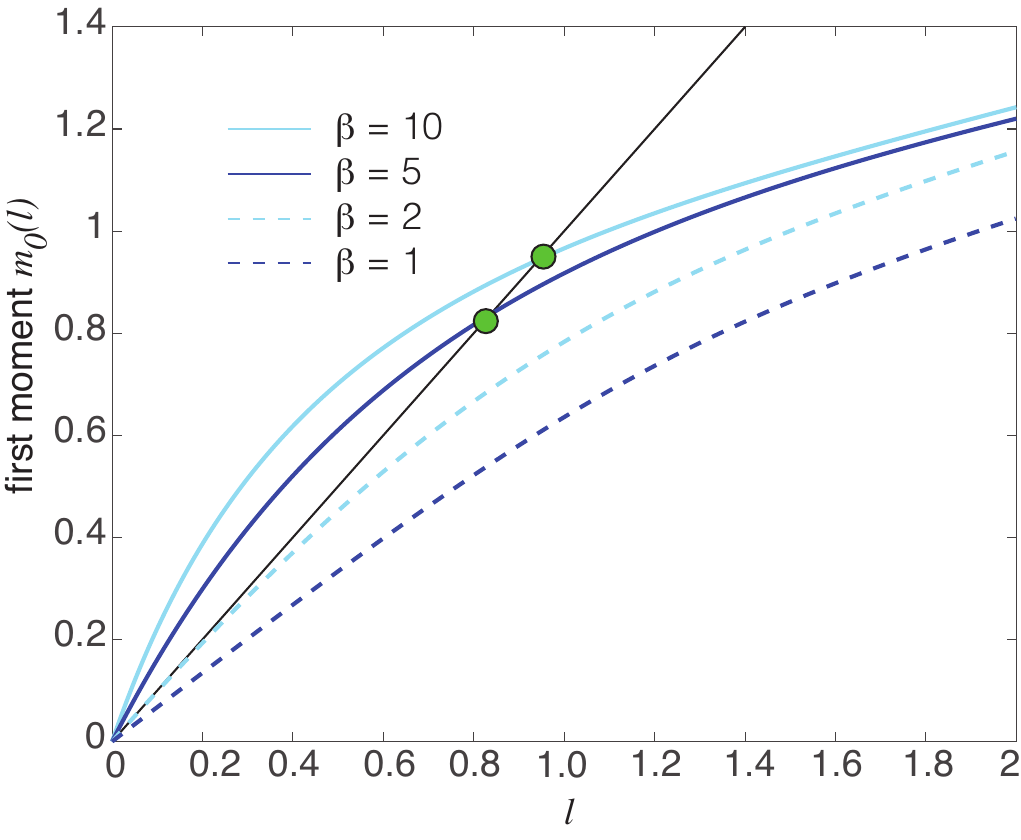} 
\caption{Stationary solution of the 1D McKean-Vlasov Eq. (\ref{MV1}) for $V(x)=x^4/4-x^2/2$ and no resetting ($r=0$). Plot of the first moment $m_0(\ell)$ as a function of $\ell$ and various inverse temperatures $\beta$. The nonzero intercepts with the diagonal determine the positive definite solution $\ell_0$. We also take $\lambda=1$.}
\label{fig1}
\end{figure}

\subsection{No resetting ($r=0$).} The steady state version of Eq. (\ref{MV1}) in the absence of resetting ($r=0$) is $J_0'(x)=0$ where
\begin{align}
 {J}_0(x)&:= -D \frac{\partial {\phi}_0(x) }{\partial x}   -\beta  D{\phi}_0(x) \bigg ( V'(x)+\lambda \int_{-\infty}^{\infty}(x-y)  {\phi}_0(y) dy\bigg ) . 
\end{align}
(We use the subscript $0$ to indicate no resetting.) 
The integral term reduces to $\lambda (x-\langle y\rangle)$ with $\langle y\rangle =\int_0^{\infty}y {\phi}_0(y)dy$. Suppose, for the moment, that $\langle y\rangle =\ell$ for some fixed $\ell$, which then acts as a parameter of the density $\phi_0$. The normalizability of $\phi_0(x)$ implies that $J_0(\pm \infty)=0$ and so $J_0(x)=0$ for all $x$. It follows that, for fixed $\ell$, the stationary density is given by a Boltzmann distribution:
\begin{equation}
\label{0ssbar}
\phi_0={\phi}_{0}(x;\ell)=Z(\ell)^{-1}\exp\left (-\beta [V(x)+\lambda x^2/2-\ell\lambda x]\right ).
\end{equation}
The factor $Z(\ell)$ ensures the normalization $\int_0^{\infty} {\phi}_{0}(x;\ell)dx=1$. 
The unknown parameter $\ell$ is determined by imposing the self-consistency condition
\begin{equation}
\label{0acon}
\ell=m_0(\ell)\equiv \int_{-\infty}^{\infty} x {\phi}_{0}(x;\ell)dx. 
\end{equation}
The number of equilibrium solutions is then equal to the number of solutions of Eq. (\ref{0acon}). First, consider the quadratic confining potential $V(x)=\nu x^2/2$, $\nu >0$. We have
\begin{align}
Z(\ell)&=\int_{-\infty}^{\infty} \e^{-\beta [(\nu+\lambda) x^2/2-\ell\lambda x]}dx=\sqrt{\frac{2\pi}{\beta[ \nu +\lambda]}}\e^{\beta \ell^2\lambda^2/2[\nu +\lambda]},
\label{Za}
\end{align}
and Eq. (\ref{0acon}) becomes
\begin{align}
 \ell &=Z(\ell)^{-1}\int_0^{\infty}x \e^{-\beta [(\nu +\lambda)x^2/2-\ell\lambda x]}dx=\frac{1}{\lambda\beta}\frac{\partial \log Z(\ell)}{\partial \ell} = \frac{ \ell \lambda}{\nu+\lambda}.
\end{align}
It follows that $\ell =0$ and
\begin{equation}
\phi_{0}(x;0)=\sqrt{\frac{\beta[ \nu +\lambda]}{2\pi}}\exp\left (-\beta (\nu+\lambda) x^2/2\right ).
\end{equation}
Hence, the interactions simply modify the effective strength of the quadratic potential.

The situation is more complicated when $V(x)$ has at least two minima, because the tendency of the Boltzmann distribution to localize around both minima competes with the cooperative effects of the Curie-Weiss potential. As an example, consider the double-well potential $V(x)=x^4/4-x^2/2$. Although it is no longer possible to analytically solve the corresponding self-consistency Eq. (\ref{0acon}), one can prove that there exists a phase transition at a critical temperature $T_c$ such that $\ell=0$ for $T>T_c$ and $\ell=\pm \ell_0\neq 0$ for $T<T_c$ \cite{Desai78,Dawson83,Pavliotis19}. This is illustrated in Fig. \ref{fig1} by plotting the first moment function $m(\ell)$ for different values of $\beta$. We find numerically that $\beta_c \approx 2$ when $\lambda=1$, which is consistent with the critical point obtained in Refs. \onlinecite{Desai78,Dawson83}.
\medskip

\subsection{With resetting ($r>0$).} The 1D steady state equation when $r>0$ is
\begin{align}
 0
&=D\frac{d^2 \phi(x)}{d x^2}-r\phi(x)+r\delta(x-x_0) +\frac{1}{\gamma}\frac{d}{d x}\bigg (\phi(x)  V_{\ell}'(x) \bigg ),
\label{MVa}
\end{align}
with
\begin{equation}
V_{\ell}(x)=V(x)+\frac{\lambda}{2}x^2-\lambda \ell x.
\end{equation}
Finding a stationary solution $\phi$ for fixed $\ell$ thus reduces to the problem of solving the time-independent diffusion equation with resetting in an effective potential $V_{\ell}(x)$.  
This is considerably more involved than the case $r=0$, since $\phi_{\ell}(x)$ now represents a nonequilibrium stationary state (NESS) rather than a Boltzmann distribution \cite{Pal15,Roldan17,Metzler20}. In other words, there exist non-zero time-independent probability fluxes, with the common reset point $x_0$ acting as a probability source, and all positions $x\neq x_0$ acting as probability sinks. 

For the sake of illustration, consider a quadratic potential $V(x)=\nu x^2/2$ so that up to an irrelevant constant, $V_{\ell}(x)$ takes the form of an harmonic potential centered at $x_{\ell}$:
\begin{equation}
V_{\ell}(x)=\frac{\lambda+\nu}{2}(x-{x}_{\ell})^2,\quad {x}_{\ell}=\frac{\ell \lambda}{\lambda+\nu}.
\end{equation}
It immediately follows that, although $V_{\ell}(x)$ is unimodal, there is competition between minimizing the energy at the point $x_{\ell}$ and resetting to the point $x_0$, assuming that $x_0\neq {x}_{\ell}$. The example of diffusion with stochastic resetting in an harmonic potential has been analyzed elsewhere \cite{Pal15}. In order to carry over previous results, we perform the translation $x\rightarrow x-{x}_{\ell}$. \pcb{Setting $\widehat{\phi}(x)=\phi(x+x_{\ell})$, Eq. (\ref{MVa}) can be rewritten as
\begin{align}
 0
&=D\frac{d^2 \widehat{\phi}(x)}{d x^2}-r\widehat{\phi}(x)+r\delta(x-x_0(\ell)) +\frac{1}{\gamma}\frac{d}{d x}\bigg (\widehat{\phi}(x)  U'(x) \bigg ),
\label{MVa2}
\end{align}
where}
\begin{equation}
U(x)=\frac{\mu x^2}{2},\quad x_0(\ell)=x_0-\frac{\ell\lambda}{\mu},\quad \mu=\lambda+\nu.
\end{equation}
All of the dependence on the self-consistent parameter $\ell$ is now in the reset point. For future reference, note that $\mu/\gamma=D\mu/k_BT=\beta D \mu$ has units of inverse time.

A solution of Eq. (\ref{MVa2}) for fixed $\ell$ can be identified with the Green's function $G(x,y)$ of the linear operator $\L$ on $\R$, where 
\begin{align}
\label{G}
 \L G(x,y)&\equiv \frac{d^2G(x,y)}{d x^2}-\frac{r}{D}G(x,y) +\beta \frac{d}{d x}\bigg (  U'(x)G(x,y) \bigg )=-\frac{r}{D}\delta(x-y) , 
\end{align}
after setting $y=x_0(\ell)$ and $\beta=1/(\gamma D)=1/(k_BT)$. 
Note that integrating Eq. (\ref{G}) with respect to $x\in [y-\epsilon,y+\epsilon$ and taking the limit $\epsilon \rightarrow 0^+$ yields
the flux discontinuity condition
\begin{equation}
\label{disc}
\left . \frac{dG(x,y)}{dx}\right |_{x=y^+}-\left . \frac{dG(x,y)}{dx}\right |_{x=y^-}=-\frac{r}{D}.
\end{equation}
Finally, we require $G(x,y)$ to be continuous at $x=y$ and to decay sufficiently fast as $x\rightarrow \pm \infty$. Not that the resulting Green's function solution  automatically satisfies the normalization condition 
$\int_{-\infty}^{\infty}G(x,y)dx=1$. 
In Ref. \onlinecite{Pal15} the NESS $\phi(x)$ was solved in terms of confluent hypergeometric or Kummer functions. (For completeness, a summary of the basic properties of confluent hypergeometric functions is given in appendix A, see also Ref. \onlinecite{Arfken85}.) Here we provide an alternative derivation based on a classical Green's function construction. 

\begin{figure}[b!]
\centering
\includegraphics[width=10cm]{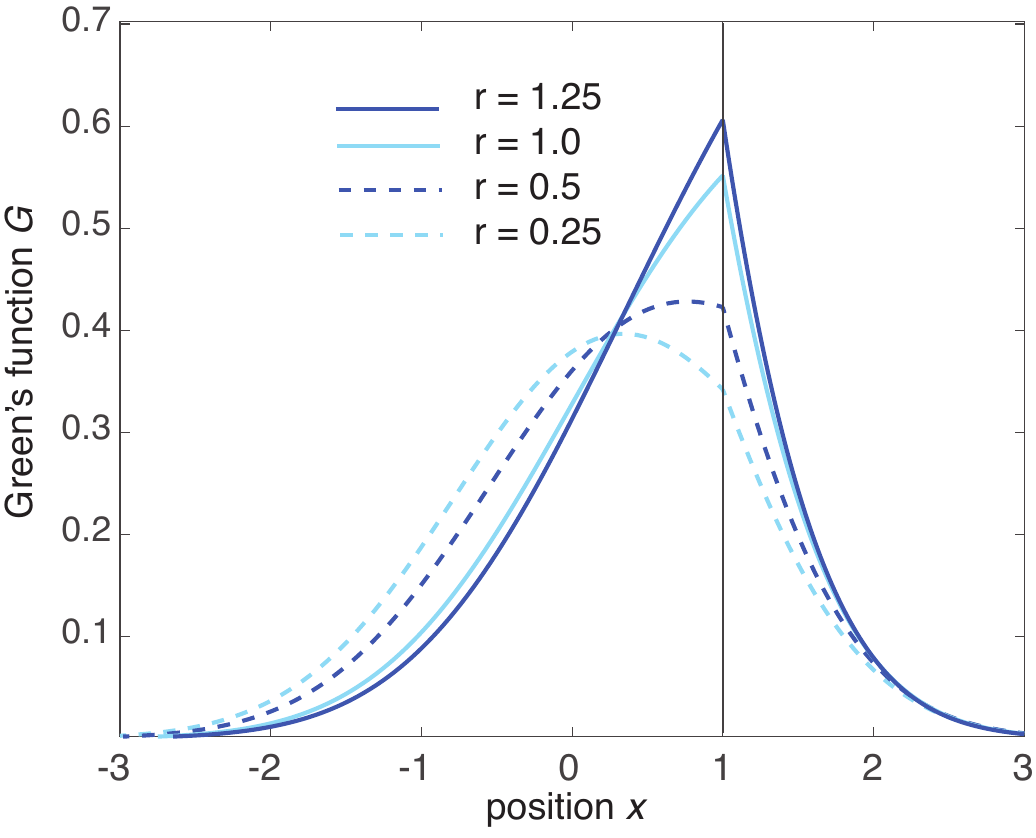} 
\caption{Plot of Green's function $G(x,y)$ satisfying Eq. (\ref{G}) for fixed $y$ and various resetting rates $r$. Other parameters are $\beta \mu =1$, $y=1$ and $D=1$.}
\label{fig2}
\end{figure}

\begin{figure}[b!]
\centering
\includegraphics[width=10cm]{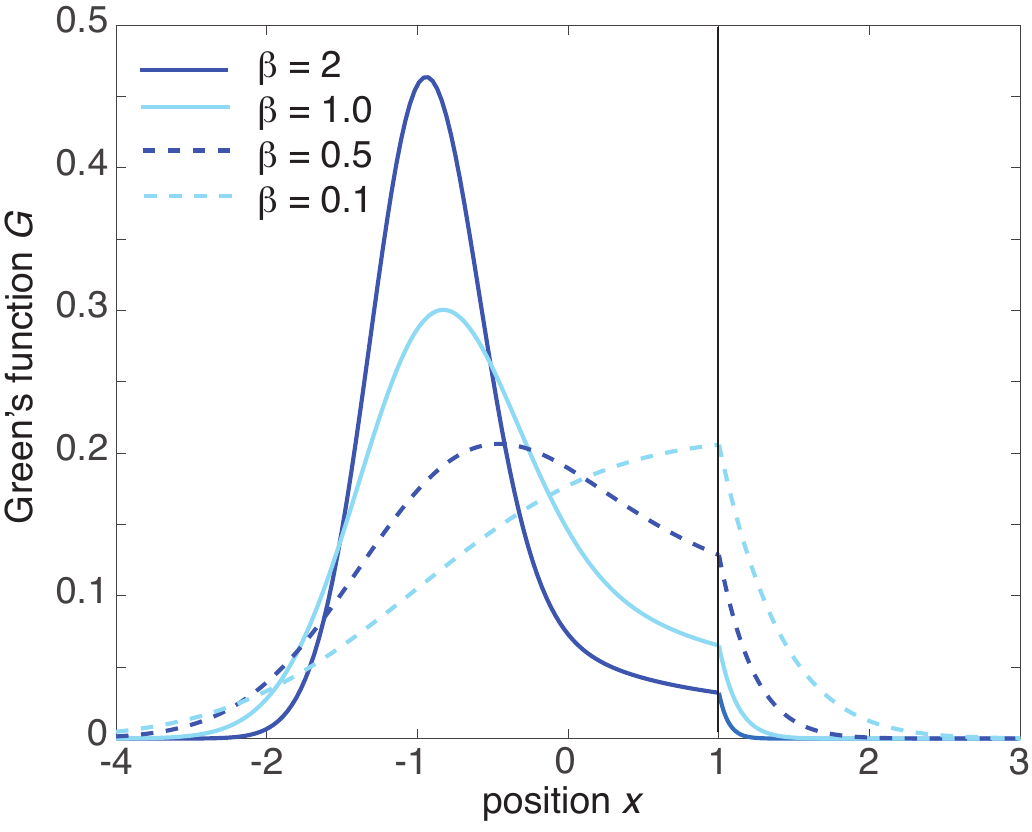} 
\caption{Plot of Green's function $G(x,y)$ satisfying Eq. (\ref{G}) for fixed $y$ and various inverse temperatures $\beta$. Other parameters are $ \mu =1$, $y=1$, $r=0.25$ and $D=1$.}
\label{fig3}
\end{figure}

First, consider the homogeneous equation $\L g(x)=0$. Performing the change of variables $g(x)=\exp(-\beta \mu x^2/2) h(x)$ we obtain the following equation for $h(x)$:
\begin{equation}
\frac{d^2h(x)}{d x^2}-\beta \mu x \frac{dh(x)}{dx}-\frac{r}{D}h(x)=0,
\end{equation}
which we rewrite in the more suggestive form
\begin{equation}
\label{h}
\frac{2}{\beta \mu}\frac{d^2h(x)}{d x^2}-2x\frac{dh(x)}{d x}-4a h(x),\quad a=\frac{r}{2D\beta \mu}.
\end{equation}
Comparison with the confluent hypergeometric Eq. (\ref{cgf2}) immediately shows that $g(x)$ has the general solution
\begin{align}
 g(x)&=A_1 \e^{-\beta \mu x^2/2}M(a,1/2,\mu \beta x^2/2)+A_2 |x|\,  \e^{-\beta \mu x^2/2}M(a+1/2,3/2,\beta \mu x^2/2),
 \label{h1}
\end{align}
where $M(a,c,x)$ is defined in Eq. (\ref{M}) and $A_{1,2}$ are constant coefficients. However, the asymptotic expansion (\ref{infM}) implies that for large $|x|$ we have $g(x)\sim x^{2a-1}$, which converges to zero algebraically when $a <1/2$ and diverges when $a>1/2$. Therefore, in order to have solutions that decays exponentially as $x\rightarrow   \infty$ we take the solution to be defined in terms of the Kummer-U function (\ref{U}):
\begin{subequations}
\begin{align}
  g_+(x) &= \e^{-\beta \mu x^2/2}U(a,1/2;\beta \mu x^2/2)\nonumber \\
 &=\sqrt{\pi}  \pcb{\e^{-\beta \mu x^2/2}}\bigg [\frac{M(a,1/2;\beta \mu x^2/2)}{(a-1/2)!} - \frac{\sqrt{2\beta \mu} x M(a+1/2,3/2;\beta \mu x^2/2)}{(a-1)!}\bigg ].
\end{align}
Similarly, a solution that decays exponentially as $x\rightarrow -\infty$ is
\begin{align}
  g_-(x) &= \e^{-\beta \mu x^2/2}\overline{U}(a,1/2;\beta \mu x^2/2)\nonumber \\
 &=\sqrt{\pi}  \pcb{\e^{-\beta \mu x^2/2}}\bigg [\frac{M(a,1/2;\beta \mu x^2/2)}{(a-1/2)!} +\frac{\sqrt{2\beta \mu} x M(a+1/2,3/2;\beta \mu x^2/2)}{(a-1)!}\bigg ].
\end{align}
\end{subequations}
We can now construct the Green's function by setting 
\begin{equation}
G(x,y)=\left \{\begin{array}{cc} Ag_+(x)g_-(y)& x>y\\Ag_+(y)g_-(x)& x<y \end{array}\right . .
\end{equation}
 $G$ clearly satisfies Eq. (\ref{G}) in the two domains $x>y$ and $x<y$,  decays exponentially fast at $x=\pm \infty$, and is continuous at $x=y$. The coefficient $A$ is determined by imposing the flux discontinuity condition (\ref{disc}). 
That is,
\begin{equation}
 A=\frac{r}{2\pi D\sqrt{2\beta \mu}}\frac{1}{f_a(y)f_{a+1/2}(y)+\beta \mu y^2h_a(y)},
\end{equation}
where $h_a(y)=[f_a(y)f'_{a+1/2}(y)-f_a'(y)f_{a+1/2}(y)]$ and
\begin{subequations}
\begin{align}
 f_a(x)&=\frac{M(a,1/2;\beta \mu x^2/2)}{(a-1/2)!}\\ f_{a+1/2}(x)&=\frac{  M(a+1/2,3/2;\beta \mu x^2/2)}{(a-1)!}.
\end{align}
\end{subequations}
In Fig. \ref{fig2} we plot the Greens's function $G(x,y)$ as a function of $x$ for various resetting rates and $y=1$ . It can clearly be seen that as $r$ increases the peak of the Green's function is more sharply defined at the reset point. On the other hand, reducing the temperature (increasing $\beta$) reduces the effects of resetting and $G(x,y)$ approaches a Boltzmann-like distribution centered at $x=0$, see Fig. \ref{fig3}.

\begin{figure}[b!]
\centering
\includegraphics[width=10cm]{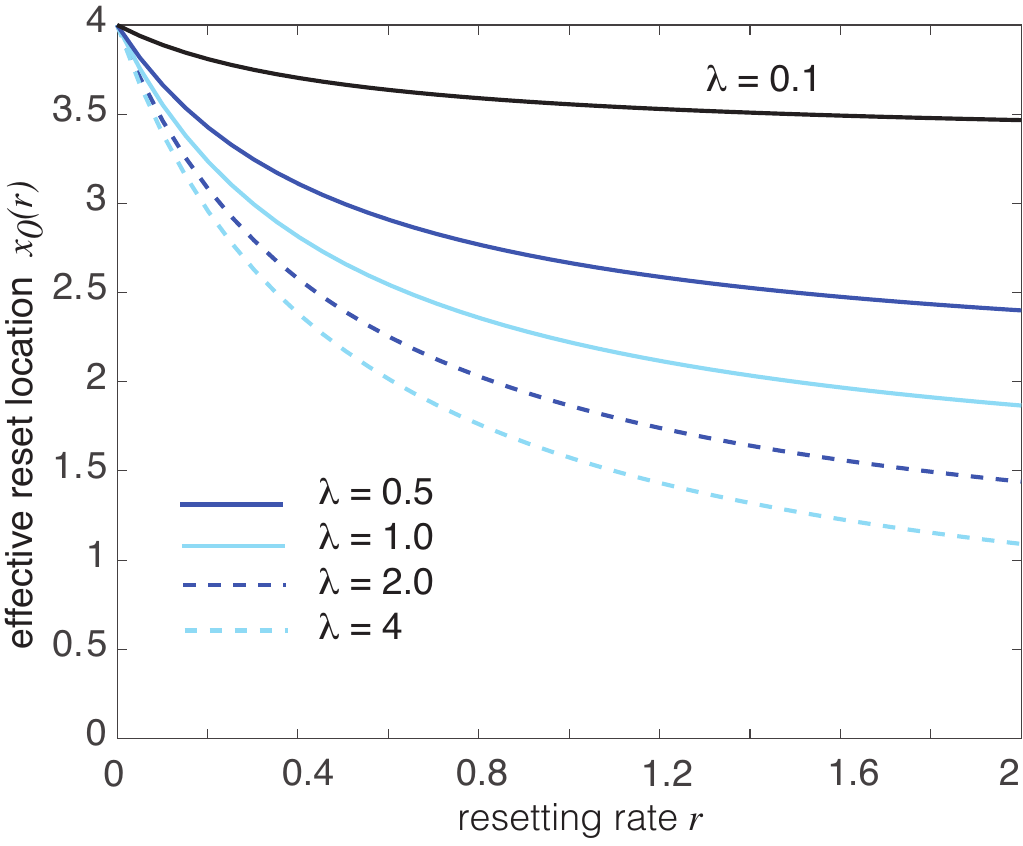} 
\caption{Solution of the1D McKean-Vlasov Eq. (\ref{MV1}) for $V(x)=\nu x^2/2$. Plot of the effective reset location $x_0(r)$, Eq. (\ref{x0r}), as a function of $r$ for various values of $\lambda$. Other parameters are $\nu =0.5$, $\beta  =1$, $D=1$, and $x_0=4$.}
\label{fig4}
\end{figure}

\pcb{The final step is to determine $\ell$ from the implicit moment Eq. (\ref{0acon}) with $\phi=\phi (x,\ell)=G(x-x_{\ell},x_0-x_{\ell})$:
\begin{equation}
\label{acon}
\ell=m(\ell)\equiv \int_{-\infty}^{\infty} (x+x_l) G(x,x_0-\lambda \ell /\mu)dx. 
\end{equation}
The function $m(\ell)$ can be calculated explicitly from the Green's function Eq. (\ref{G}).} Multiplying both sides by $x$ and integrating with respect to $x$ using integration by parts gives
\begin{equation}
\left [\beta \mu+\frac{r}{D}\right ] \int_{-\infty}^{\infty} x G(x,y)dx =\frac{ry}{D}.
\end{equation}
Hence,
\begin{equation}
\pcb{m(\ell)=x_{\ell}+\frac{r}{D}\frac{x_0-x_{\ell}}{\beta \mu+r/D}.}
\end{equation}
Substituting into Eq. (\ref{acon}) and rearranging, \pcb{we find that $\ell$ is independent of the interaction strength $\lambda$:
\begin{equation}
\label{ellr}
\ell=\ell(r)\equiv \left [\frac{r/D}{\beta \nu+r/D}\right ]x_0.
\end{equation}
A number of results follow from this analysis.}
\medskip

\begin{figure}[t!]
\centering
\includegraphics[width=10cm]{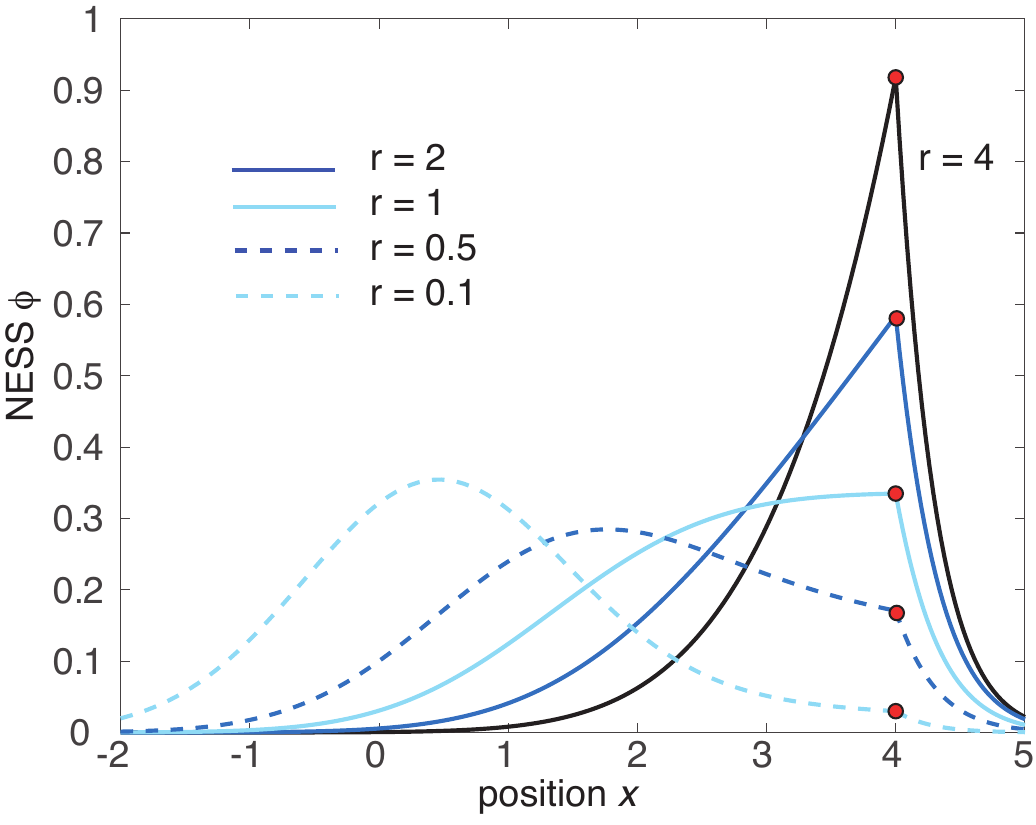} 
\caption{Solution of the1D McKean-Vlasov Eq. (\ref{MV1}) for $V(x)=\nu x^2/2$ and $r>0$. Plot of the NESS $\phi(x)$ as a function of $x$ for various resetting rates $r$ and weak coupling $\lambda=0.5$. Other parameters are $ \nu =0.5$, $\beta  =1$, $D=1$, and $x_0=4$. Filled circles indicate $x_0$.}
\label{fig5}
\end{figure}

\begin{figure}[t!]
\centering
\includegraphics[width=10cm]{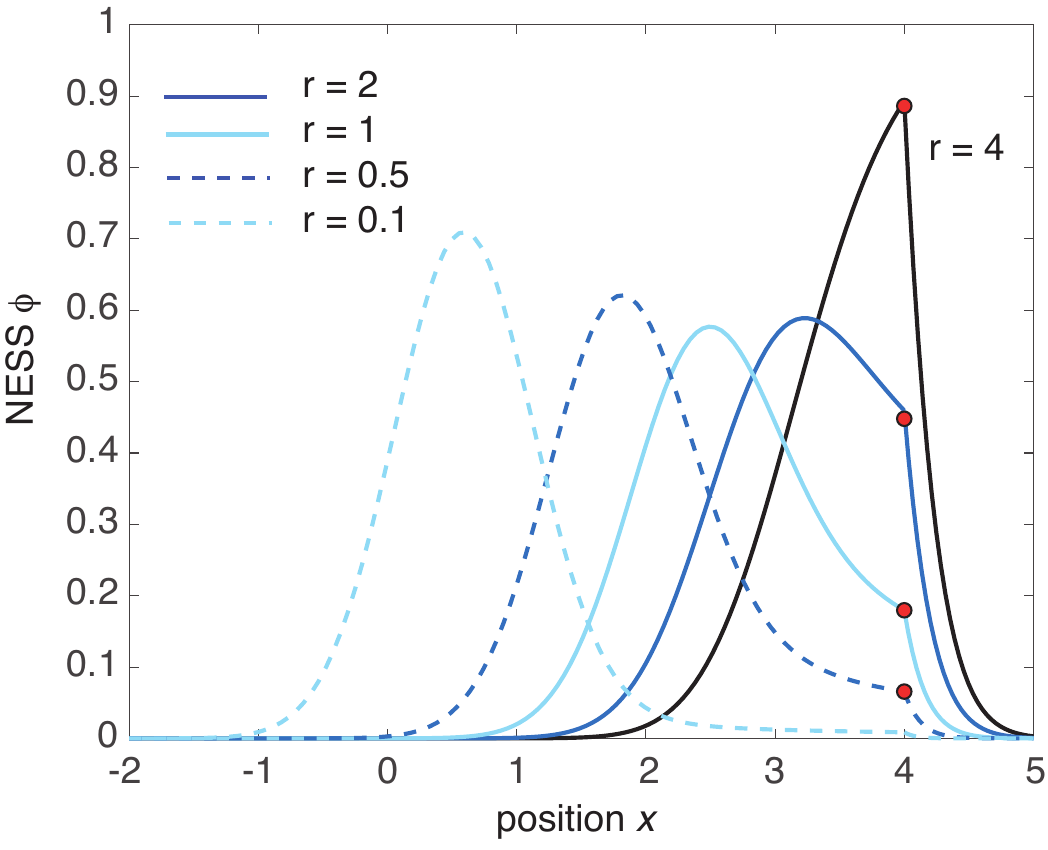} 
\caption{Solution of the1D McKean-Vlasov Eq. (\ref{MV1}) for $V(x)=\nu x^2/2$ and $r>0$. Plot of the NESS $\phi(x)$ as a function of $x$ for various resetting rates $r$ and strong coupling \pcb{$\lambda=3$. Other parameters are $ \nu =0.5$}, $\beta  =1$, and $x_0=4$. Filled circles indicate the solution at $x_0$.}
\label{fig6}
\end{figure}

\noindent (i) There exists a unique self-consistent stationary solution whose first moment is $\ell(r)$ for a fixed resetting rate $r$, Eq. (\ref{ellr}) implies that 
 that the effective resetting point in Eq. (\ref{MVa2}) is
\begin{equation}
x_0(r) :=x_0 -\frac{\lambda \ell(r)}{\mu}=x_0\left [1-\frac{\lambda}{\mu}\frac{r/D}{\beta \nu+r/D}\right ].
\label{x0r}
\end{equation}
Example plots of $x_0(r)$ are shown in Fig. \ref{fig4}. Note that 
\begin{equation}
\lim_{r\rightarrow \infty}x_0(r)=x_0^*=\frac{\nu x_0}{\lambda+\nu}.
\end{equation}
\medskip

\noindent (ii) For small $r$, the NESS is dominated by the effective potential $V_{0}(x)$ and $\phi(x)$ is approximately equal to a unimodal function centered around $x=0$. As $r$ increases, stochastic resetting becomes more dominant and $\phi(x)$ shifts rightward towards $x_0$. For sufficiently, large $r$, the NESS is sharply localized around $x=x_0$. The changes in $\phi(x)$ as $r$ increases are illustrated in Figs. \ref{fig5} and Fig. \ref{fig6} for small and large coupling, respectively.
\medskip

\noindent (iii) \pcb{Although the mean $\ell=\langle x\rangle $ is independent of the coupling strength $\lambda$, higher order moments are $\lambda$-dependent.}
\setcounter{equation}{0}

\section{Statistical correlations and global resetting}

In the case of global resetting the MV Eq. (\ref{MVglob}) is driven by an external shot noise process, which induces statistical correlations. That is, 
\begin{equation}
\E[\Phi(\x,t)\Phi((\y,t)]\neq \E[\Phi(\x,t)]\E[\Phi(\x,t)],
\end{equation}
and similarly for higher-order moments. In order to explore this issue in more detail we ignore particle interactions so that the MV Eq. (\ref{MVglob}) reduces to the linear stochastic FP equation 
\begin{align}
  \frac{\partial \Phi(\x,t)}{\partial t} 
&=D{\bm \nabla}^2  \Phi(\x,t)-{\bm \nabla} \cdot \bigg (\Phi(\x,t) {\bf A}(\x)\bigg ) +[\rho_0(\x)-\Phi(\x,t)]h(t).
\label{FPglob}
\end{align}
For convenience, we have set ${\bf A}(\x)=-\gamma^{-1} {\bm \nabla}V(\x)$.

Following our previous work on actively switching (hybrid) systems \cite{Bressloff15a,Bressloff15b,Bressloff16}, we spatially discretize Eq. (\ref{FPglob}) in terms of a $d$-dimensional regular lattice with nodes ${\bm \ell} \in \Z^d$ and lattice spacing $s$. Let $\Gamma_{{\bm \ell}}=\{{\bm \ell} \pm s{\bf e}_{\sigma}, \sigma=1,\ldots d\}$ denote the set of nearest neighbors of ${\bm \ell}$, 
where ${\bf e}_{\sigma}$ is the unit vector along the $\sigma$-axis.
Setting $U_{{\bm \ell}}(t)=\Phi(s{\bm \ell} ,t)$, $u_{\ell,0}=\rho_0(s{\bm \ell})$ and $A_{{\bm \ell}}^{\sigma}=A^{\sigma}({{\bm \ell}} s)$, ${\bm \ell}\in {\mathbb Z}^d$, we obtain a piecewise deterministic ODE with stochastic resetting
\begin{align}
\label{sh}
\frac{du_{\bm \ell}(t)}{dt}&=D[\Delta(\u(t))]_{\bm \ell} -\frac{1}{s}\sum_{\sigma=1}^d [u_{{\bm \ell}+{\bf e}_{\sigma}}(t)A^{\sigma}_{{\bm \ell}+{\bf e}_{\sigma} }-u_{\bm \ell}(t)A_{\bm \ell}^{\sigma}] +[u_{{\bm \ell},0}-u_{\bm \ell}(t)]h(t),
\end{align}
where $\Delta$ is the discrete Laplacian 
\begin{equation}
\label{dL1}
[\Delta(\u)]_{{\bm \ell}}=\frac{1}{h^2} \sum_{{\bm \ell}'\in \Gamma_{{\bm \ell}}}[u_{{\bm \ell}'}-u_{{\bm \ell}}] .
\end{equation}
Introducing the infinite-dimensional vector $\U=(u_{{\bm \ell}},\, {\bm \ell} \in \Z^d)$,  and the corresponding probability density
${\mathcal P}(\u,t)d\u= \P\{\U(t)\in [\u,\u+d\u]\}$,
 we write down the following modified Liouville equation for the spatially discretized system: 
\begin{align}
  \frac{\partial \calP}{\partial t}&=-\sum_{{\bm \ell}\in \Z^d}\frac{\partial}{\partial u_{{\bm \ell}}}\bigg [\bigg (D[\Delta(\u)]_{\bm \ell}  -\frac{1}{s}\sum_{\sigma=1}^d[u_{{\bm \ell}+{\bf e}_{\sigma}}A^{\sigma}_{{\bm \ell}+{\bf e}_{\sigma} }-u_{{\bm \ell}}A_{{\bm \ell}}^{\sigma}]\bigg )\calP(\u,t)\bigg ] \nonumber \\
 &\quad -r\calP(\u,t)+r\delta (\u-\u_0).
\label{swCK00}
\end{align}
Finally, we take the continuum limit $s\rightarrow 0$ of Eq. (\ref{swCK00}). This  
yields a functional Liouville equation for the many-body probability functional $\calP[u,t]$ with
\begin{equation}
\calP[u,t]\prod_{\x}du(\x)= \lim_{s\rightarrow 0}\calP_n({\bf u},t)d\u.
\end{equation}
That is,
\begin{align}
 \frac{\partial \calP_n[u,t]}{\partial t} &=-\int_{\R^d} d\x\, \frac{\delta}{\delta u(\x)}\bigg [ \calP[u,t]  \bigg (D{\bm \nabla}^2u(\x)\ -{\bm \nabla}\cdot [u(\x){\bf A}(\x)]\bigg )\bigg ] -r\calP[u,t] +r\calP_0[u],
 \label{swCK0}
\end{align}
where $\calP_0[u]=\prod_{\x}\delta(u(\x)-u_0(\x))$.

Eq. (\ref{swCK0}) can now be used to derive various moment equations with respect to the shot noise. For example, the first moment is defined as
\begin{align}
\label{mfirst}
{\mathcal U}(\x,t)&\equiv  \E[u(\x,t) 1_{N(t)=n}]=\int {\mathcal D}[u]\, u(\x)\calP[u,t].
\end{align}
Multiplying both sides of Eq. (\ref{swCK0}) by $u(\x)$ and functionally integrating with respect to $u$ gives
\begin{align*}
  &\frac{\partial }{\partial t}\int {\mathcal D}[u] u(\x)\calP_n[u,t]\nonumber \\
 &=- \int {\mathcal D}[u] u(\x)\bigg \{\int_{\R^d} d\x'\, \frac{\delta}{\delta u(\x')} \bigg [\calP[u,t]\bigg ( (D{\bm \nabla}^2u(\x')-{\bm \nabla}\cdot [u(\x'){\bf A}_n(\x')]\bigg ) \bigg ]\nonumber \\
   &\quad -r \calP[u,t]+r\calP_0[u]\bigg \}.
\end{align*}
Functionally integrating by parts and using the functional derivative identity $\delta u(\x)/\delta u(\x')=\delta(\x-\x')$, we have
\begin{align*}
 \frac{\partial }{\partial t}\int {\mathcal D}[u] u(\x)\calP[u,t]&=\int {\mathcal D}[u] \bigg \{ \bigg [\left (D{\bm \nabla}^2u(\x)-{\bm \nabla}\cdot [u(\x){\bf A}(\x)]\right ) \calP[u,t]\bigg ]\nonumber \\
   &\quad -ru(\x)\calP[u,t]+ru(\x)\calP[u,0]\bigg \},
\end{align*}
which is equivalent to the first moment equation 
\begin{align}
\label{mom1}
   \frac{\partial {\mathcal U(\x,t)}}{\partial t}&= D{\bm \nabla}^2{\mathcal U}(\x,t)-{\bm \nabla}\cdot [ {\mathcal U}(\x,t){\bf A}(\x) ] -r{\mathcal U}(\x,t)+r{\mathcal U}_0(\x).
\end{align}
Note that Eq. (\ref{mom1}) is identical in form to the deterministic FP equation for non-interacting Brownian particles with local resetting, see Eq. (\ref{MVloc}).
On the other hand, the second-order moments
\begin{equation}
C(\x,\y,t)=\E[u(\x,t)u(\y,t)]
\end{equation}
evolve according to the equation
  \begin{align}
   \frac{\partial C}{\partial t} &=D{\bm \nabla}_{\x}^2 C(\x,\y,t)+D{\bm \nabla}_{\y}^2C(\x,\y,t)   -{\bm \nabla}_{\x}\cdot [C(\x,\y,t){\bf A}(\x)] -{\bm \nabla}_{\y}\cdot [C(\x,\y,t){\bf A}(\y) ]\nonumber \\
 &\quad  -rC(\x,\y,t)+rC_0(\x,\y). 
 \label{C22} 
  \end{align}
The latter can be derived from Eq. (\ref{swCK0}) after multiplying both sides by the product $u(\x)u(\y)$ and functionally integrating by parts along similar lines to the first moments. 

One of the major implications of the moment equations is that global resetting introduces statistical correlations between the particles, even in the absence of particle interactions. For example, the two-point correlation function is non-zero since $C(\x,\y,t)\neq \calU(\x,t)\calU(\y,t)$. We establish the latter using proof by contradiction. Suppose that $C(\x,\y,t)= \calU(\x,t)\calU(\y,t)$. Substituting this trial solution into Eq. (\ref{C22}) gives
 \begin{align}
    \calU(\y,t)\frac{\partial  \calU(\x,t)}{\partial t}+\calU(\x,t)\frac{\partial  \calU(\y,t)}{\partial t} &=D\calU(\y,t){\bm \nabla}_{\x}^2 \calU(\x,t)+D\calU(\x,t){\bm \nabla}_{\y}^2\calU(\y,t)\nonumber \\
   &- \calU(\y,t){\bm \nabla}_{\x}\cdot [\calU_n(\x,t){\bf A}(\x)]-\calU(\x,t){\bm \nabla}_{\y}\cdot [\calU(\y,t){\bf A}(\y) ]\nonumber \\
   &  \quad 
  -r\calU(\x,t)\calU(\y,t)+r\calU_0(\x)\calU_0(\y). 
  \end{align}
  Applying the first moment Eq. (\ref{mom1}) then implies that
   \begin{align*}
    0 &=[\calU(\x,t)-\calU_0(\x)]\calU(\y,t)+[\calU(\y,t)-\calU_0(\y)]\calU(\x,t) -[\calU(\x,t)\calU(\y,t)-\calU_0(\x)\calU_0(\y)] \nonumber \\
    &=\calU(\x,t)\calU(\y,t)+\calU_0(\x)\calU_0(\y)-\calU_0(\x)\calU(\y,t) -\calU_0(\y)\calU(\x,t),
  \end{align*}
 which is clearly false. Hence, $C(\x,\y,t)\neq \calU(\x,t)\calU(\y,t)$. Similar results hold for higher-order moments.

 
\setcounter{equation}{0}

\section{Kuramoto model with stochastic resetting}

One of the most studied interacting particle systems is the Kuramoto model of weakly-coupled,
near identical limit-cycle oscillators with a sinusoidal phase interaction function \cite{Kuramoto84,Strogatz00,Acebron05}. The deterministic version of the model takes the form of a system of nonlinear phase equations
\begin{equation}
\label{Kur1}
\frac{d\theta_j}{dt}=\omega_j+\frac{\lambda}{N}\sum_{k=1}^N \sin(\theta_k-\theta_j),
\end{equation}
where $\theta_j(t)\in [0,2\pi]$ is the phase of the $j$th oscillator with natural frequency $\omega_j$, and $\lambda \geq 0$ is the coupling strength. The frequencies $\omega_j$ are typically assumed to be distributed according to a probability density $g(\omega)$ with (i) $g(-\omega)=g(\omega)$
and (ii) $g(0) \geq g(\omega)$ for all $w \in [0,\infty)$. Without loss of generality, one can always take $g(\omega)$ to have
zero mean by going to a rotating frame if necessary. One method for investigating the collective behavior of the Kuramoto model is to assume that it has a well-defined mean field limit $N\rightarrow \infty$ involving a continuum of oscillators distributed
on the circle \cite{Strogatz91,Crawford94,Crawford99}. Let
$\sigma_0(\theta,t,\omega)$ denote the fraction of oscillators with natural frequency $\omega$ that lie between $\theta$ and $\theta + d\theta$
at time $t$ with
\begin{equation}
\int_0^{2\pi}\sigma_0(\theta,t,\omega)d\theta =1. 
\end{equation}
More precisely, $\sigma_0(\theta,t,\omega)$ is a population density that is conditioned on the natural frequency of the oscillators, see below.
Since the total number of oscillators is fixed, $\sigma$ evolves according to the continuity or Liouville equation
\begin{equation} 
\label{Kurd}
\frac{\partial \sigma_0}{\partial t}=-\frac{\partial}{\partial \theta}(\sigma_0 v_0),
\end{equation}
where
\begin{align}
\label{vK}
v_0(\theta,t,\omega)=\omega + \lambda \int_{0}^{2\pi}d\theta' \, \int_{-\infty}^{\infty}d\omega'
\sin(\theta'-\theta)\sigma_0(\theta',t,\omega')g(\omega'). 
\end{align}
It is also possible to consider a stochastic version of the Kuramoto model \cite{Sakaguchi88}. If $\Theta_j(t)\in [0,2\pi)$ denotes the stochastic phase of the $j$th oscillator at time $t$, then the corresponding SDE 
 is 
\begin{equation}
 d\Theta_j(t)=\bigg [\omega_j+\frac{\lambda}{N}\sum_{k=1}^N \sin(\Theta_k(t)-\Theta_j(t))\bigg ]dt+\sqrt{2D}dW_j(t)\label{Kur2}
\end{equation}
for $  j=1,\ldots,N$, where $W_j(t)$ is an independent Wiener process. The corresponding continuum model now takes the form of a nonlinear FP equation on the circle:
\begin{equation} 
\frac{\partial \sigma_0}{\partial t}=-\frac{\partial}{\partial \theta}(\sigma_0 v_0)+D\frac{\partial^2\sigma_0}{\partial \theta^2}.
\label{Kur3}
\end{equation}

An alternative interpretation of the SDE (\ref{Kur2}) is a system of Brownian particles on an $N$-torus with pairwise coupling and quenched disorder due to the random distribution of natural frequencies. It follows that Eq. (\ref{Kur3}) is equivalent to  the corresponding MV equation for the global density in the mean field limit. The existence of the latter has been proven rigorously using propagation of chaos \cite{Dai96}, and has been extended to a wider class of interacting particle systems on the torus \cite{Carrillo20,Pavliotis21}. The mean field limit also applies to the deterministic Kuramoto model in the case of an ensemble of initial conditions. (Alternatively, Eq. (\ref{Kurd}) can be derived within a fully deterministic framework \cite{Neunzert78,Lancellotti05}.) In this section we exploit the connection between the stochastic Kuramoto model and systems of interacting Brownian particles on a torus to incorporate the effects of stochastic resetting along analogous lines to Sec. II.

\subsection{Dean-Kawasaki equation}

Suppose that the $j$th oscillator resets to its initial phase $\theta_{j,0}$ at the times $T_{j,n}$, which are generated from a Poisson process at a rate $r$. In the case of local resetting, the reset times of the oscillators are statistically independent, whereas $T_{j,n}=T_n$ for all $j=1,\ldots,N$ when resetting is global. Analogous to the SDE for an overdamped Brownian gas, see Eq. (\ref{SDEres}), the stochastic Kuramoto model takes the form
\begin{align}
 \frac{d\Theta_j(t)}{dt}&=\omega_j+\sqrt{2D}\xi_j(t)+\frac{\lambda}{N}\sum_{k=1}^N\sin(\Theta_k(t)-\Theta_j(t)) +\sum_{n=1}^{\infty}(\theta_{j,0}-\Theta_j(t))\delta(t-T_{j,n}).
 \label{SDEKur}
\end{align}
Introduce the global density or empirical measure
\begin{equation}
\rho(\theta,t,\omega)=\frac{1}{N}\sum_{j=1}^N\delta (\theta-\Theta_j(t))\delta(\omega-\omega_j).
\end{equation}
It is important to note that the stochastic density $\rho$ is distinct from the deterministic density $\sigma_0$ for the noiseless Kuramoto model. Moreover, the former has the normalization
\begin{equation}
\int_0^{2\pi} \rho(\theta,t,\omega)d\theta =\frac{1}{N}\sum_{j=1}^N\delta(\omega-\omega_j).
\end{equation}
Taking expectations with respect to the quenched random frequencies, we have
\begin{align}
\E[\rho(\theta,t,\omega)]&=\frac{1}{N}\sum_{j=1}^N\delta (\theta-\Theta_j(t))\E[\delta(\omega-\omega_j)]=\frac{g(\omega)}{N}\sum_{j=1}^N\delta (\theta-\Theta_j(t)),
\end{align}
which implies that
$\int_0^{2\pi} \E[\rho(\theta,t,\omega)]d\theta =g(\omega)$.

Consider an arbitrary smooth test function $f: [0,2\pi]\times\R\rightarrow \R$.
Using It\^o's lemma to Taylor expand $f(\Theta_j(t+dt),\omega_j)$ and setting 
\begin{equation*}
\pcb{N^{-1}\sum_{j=1}^N f(\Theta_j(t),\omega_j)=\int_0^{2\pi}d\theta \, \int_{\R}d\omega  \rho(\theta,t,\omega)f(\theta,\omega),}
\end{equation*}
 \pcb{we have
\begin{align*}
 &\int_0^{2\pi}d\theta \, \int_{\R}d\omega  f(\theta,\omega)\frac{\partial \rho(\theta,t,\omega)}{\partial t}  =\lim_{\Delta t\rightarrow 0}\frac{1}{N}\sum_{j=1}^N \frac{f(\Theta_j(t+\Delta t),\omega_j)-f(\Theta_j(t),\omega_j)}{\Delta t} ,
 \end{align*}
 with}
 \begin{align*}
&\Delta f  =   \partial_{\theta}f(\Theta_j(t),\omega_j)\Delta \Theta_j(t)+ \partial_{\theta}^2f(\Theta_j(t),\omega_j)\Delta \Theta_j(t)^2+\ldots \nonumber \\
 & =  \partial_{\theta}f(\Theta_j(t),\omega_j) \bigg [\omega_j+\frac{\lambda}{N}\sum_{k=1}^N \sin(\Theta_k(t)-\Theta_j(t))\bigg ]\Delta t\nonumber \\&\quad +\sqrt{2D}\partial_{\theta}f(\Theta_j(t),\omega_j)\Delta W_j(t) + D\partial_{\theta}^2f(\Theta_j(t),\omega_j)\Delta t \nonumber \\
 &\quad +   \bigg [f(\theta_{0,j},\omega_j)-f(\Theta_j(t),\omega_j)\bigg ]\sum_{n\geq 1}  \delta_{t,T_{j,n}}+o(\Delta t).
\end{align*}
\pcb{Taking the limit $\Delta t \rightarrow 0$ and using the definition of $\rho$ gives
\begin{align*}
 &\int_0^{2\pi}d\theta \, \int_{\R}d\omega f(\theta,\omega)\frac{\partial \rho(\theta,t,\omega)}{\partial t}
=\int_0^{2\pi}d\theta \,\int_{\R}d\omega\, \bigg [ \partial_{\theta} f(\theta,\omega)  \frac{\sqrt{2D}}{N}\sum_{j=1}^N \rho_j(\theta,t,\omega)\xi_j(t)\nonumber \\
&\quad  +\rho(\theta,t,\omega)\bigg (D\partial_{\theta}^2 f(\theta,\omega)   +\partial_{\theta} f(\theta,\omega)\calV[\theta,t,\omega,\rho]\bigg )\nonumber\\
 &\quad +\frac{1}{N}\sum_{j=1}^Nh_j(t)  [\delta(\theta-\theta_{0,j})\delta(\omega-\omega_j)  -\rho_j(\theta,t,\omega)]f(\theta,\omega)\bigg ],
\end{align*}
where $h_j(t)$ is the Poisson shot noise process (\ref{shot}) and
with
\begin{equation}
\calV[\theta,t,\omega,\rho]=\omega+\lambda \int_{0}^{2\pi}d\theta' \,\int_{\R}d\omega '\,   \rho(\theta',t,\omega')\sin(\theta'-\theta) .
\end{equation}
Integrating by parts the various terms involving derivatives of $f$ and using the fact that $f$ is arbitrary yields the following SPDE for $\rho$}:
\begin{align}
 \frac{\partial \rho(\theta,t,\omega)}{\partial t}&=-\sqrt{\frac{2D}{N^2}}\sum_{j=1}^{N}\frac{\partial}{\partial \theta}\bigg [ \rho_j(\theta,t,\omega) { \xi}_j(t)\bigg ]+D\frac{\partial^2}{\partial \theta^2}\rho(\theta,t,\omega)\nonumber \\
 &\quad- \frac{\partial}{\partial \theta} \bigg (\rho(\theta,t,\omega) \calV[\theta,t,\omega,\rho]\bigg )\nonumber \\
 & \quad +\frac{1}{N}\sum_{j=1}^N [\delta(\theta-\theta_{0,j})\delta(\omega-\omega_j)-\rho_j(\theta,t,\omega)]h_j(t),\label{rhoKur}\end{align}

Following along similar lines to the derivation of Eq. (\ref{rhoc}), we introduce the white noise term
\begin{equation}
{ \xi}(\theta,t,\omega)=-\frac{1}{N}\sum_{j=1}^{N}\partial_{\theta}  \bigg [ \rho_j(\theta,t,\omega)\xi_j(t)\bigg ],
\end{equation}
which has zero mean and correlation function
\begin{align*}
 &\langle 
\xi (\theta,t,\omega)\xi (\theta',t',\omega')\rangle= \frac{1}{N^2}\delta(t-t')\sum_{j=1}^{N} \partial_{\theta}\partial_{\theta'} \bigg (\rho_j(\theta,t,\omega) \rho_j(\theta',t,\omega') \bigg ).
\end{align*}
Since $\rho_j(\theta,t,\omega) \rho_j(\theta',t,\omega') =\delta(\theta-\theta')\delta(\omega-\omega')\rho_j(\theta,t,\omega)$, it follows that
\begin{align*}
&\langle 
 \xi (\theta,t,\omega)\xi (\theta',t',\omega')\rangle= \frac{1}{N}\delta(t-t')  \partial_{\theta}\partial_{\theta'} \bigg (\delta(\theta-\theta')\delta(\omega-\omega')\rho(\theta,t,\omega)  \bigg ).\nonumber
\end{align*}
Finally, we introduce the global density-dependent noise field
\begin{equation}
\widehat{\xi }(\theta,t,\omega)=\frac{1}{\sqrt{N}}\partial_{\theta} \bigg ({\eta}(\theta,t,\omega)\sqrt{\rho(\theta,t,\omega)}\bigg ),
\end{equation}
where ${\bf \eta}(\theta,t,\omega)$ is a global white noise field whose components satisfy 
\begin{equation}
\langle  \eta (\theta,t,\omega)\eta(\theta',t',\omega')\rangle =\delta(t-t')\delta(\theta-\theta)\delta(\omega-\omega').
\end{equation}
It can be checked that the Gaussian noises ${ \xi}$ and $\widehat{\xi}$ have the same correlation functions and are thus statistically identical. We thus obtain the generalized DK equation for the stochastic Kuramoto model with resetting:
\begin{align}
 \frac{\partial \rho(\theta,t,\omega)}{\partial t} &=\sqrt{\frac{2D}{N}} \frac{\partial}{\partial \theta}\bigg [ \sqrt{\rho(\theta,t,\omega)} \eta(\theta,t,\omega)\bigg ]+D\frac{\partial^2}{\partial \theta^2} \rho(\theta,t,\omega) \nonumber \\
 &\quad - \frac{\partial}{\partial \theta} \bigg (\rho(\theta,t,\omega)\calV[\theta,t,\omega,\rho]\  \bigg )\label{rhoKurc} \\
 &\quad +\frac{1}{N}\sum_{j=1}^N [\delta(\theta-\theta_{0,j})\delta(\omega-\omega_j)-\rho_j(\theta,t,\omega)]h_j(t).\nonumber
\end{align}

As in the case of the DK Eq. (\ref{rhoc}), taking expectations with respect to the white noise and resetting processes results in a moment closure problem. In the absence of resetting ($h_j=0$), it can be proven that $ \rho(\theta, t,\omega) $ converges in distribution to the solution $\phi(\theta,t,\omega)$ of an MV equation under the normalization \cite{Dai96}
\begin{equation}
\int_0^{2\pi}\phi(\theta,t,\omega)d\theta = g(\omega).
\label{norm}
\end{equation}
In order to incorporate the effects of resetting, we will proceed along analogous lines to Sec. II by introducing a mean field ansatz. Again the details will depend on whether the resetting is local or global. 

\subsection{McKean-Vlasov equation: local resetting} Taking expectations of Eq. (\ref{rhoKurc}) with resect to all of the noise processes along identical lines to the derivation of Eq. (\ref{MVloc}),
leads to the following deterministic MV equation for the mean field $\phi(\theta,t,\omega)=\bigg \langle \E[\rho(\theta,t,\omega)]\bigg \rangle$:
\begin{align}
 \frac{\partial \phi(\theta,t,\omega)}{\partial t} 
&=D\frac{\partial^2}{\partial \theta^2} \phi(\theta,t,\omega)  -\frac{\partial}{\partial \theta} \bigg (\phi(\theta,t,\omega) \calV[\theta,t,\omega,\phi] \bigg ) -r\phi(\theta,t,\omega)+r \rho_0(\theta,\omega),
\label{MVKur}\end{align}
with
\begin{equation}
\rho_0(\theta,\omega)=\frac{1}{N}\sum_{j=1}^N\delta(\theta-\theta_{j,0})\delta(\omega-\omega_j).
\end{equation}
We have imposed a mean field ansatz and used Campbell's theorem for the Poisson shot noise processes.

As in the case of interacting Brownian particles with resetting, see Sec. II, a stationary solution of the MV equation (\ref{MVKur}) takes the form of an NESS that has to be determined self-consistently. Now, however, the self-consistency condition involves the first circular moment
\begin{equation}
\label{op}
 Z_1(t)=R(t)\e^{i\psi(t)}:=\int_{0}^{2\pi}d\theta \, \int_{-\infty}^{\infty}d\omega\ \e^{i\theta}\phi(\theta,t,\omega),
\end{equation}
rather than the first moment $\ell=\langle x\rangle$ for the Curie-Weiss potential, say. Substituting Eq. (\ref{op}) into the MV Eq. (\ref{MVKur}) gives
\begin{align} 
 \frac{\partial \phi(\theta,t,\omega)}{\partial t}&=D \frac{\partial^2 \phi(\theta,t,\omega)}{\partial \theta^2}- \frac{\partial}{\partial \theta} \bigg [ \bigg (\omega + \lambda R(t)\sin(\psi(t)-\theta(t))\bigg )\phi(\theta,t,\omega) \bigg ]\nonumber \\
 & -r\phi(\theta,t,\omega)+r \rho_0(\theta,\omega).
\label{MVKur0}
\end{align}
The amplitude $R(t)$ is a measure of the degree of synchronization with $R=1$ signifying complete synchrony and $R=0$ corresponding to the incoherent state $\phi(\theta,t,\omega)=g(\omega)/2\pi$, which is a solution of Eq. (\ref{MVKur}). In principle, one could now proceed by solving the time-independent version of (\ref{MVKur0}) for fixed $Z_1$ and then substituting the resulting stationary solution $\phi_{Z_1}(\theta, \omega)$ into equation (\ref{op}) to determine $Z_1$. However, the calculation of $\phi_{Z_1}$ is nontrivial even when $r=0$, since we no longer have a stationary Boltzmann distribution on the circle. 

An alternative representation of the MV equation is obtained by considering the Fourier series expansion
\begin{eqnarray}
\phi(\theta,t,\omega)=\frac{g(\omega)}{2\pi}\left (1+\sum_{m=1}^{\infty}\left [\phi_m(\omega,t)\e^{im\theta}+\mbox{c. c.}\right ]\right ),\nonumber \\
\label{FT}
\end{eqnarray}
with
\begin{equation}
\phi_m(\omega,t)=\langle \e^{-im\theta}\rangle :=\int_0^{2\pi}\e^{-im\theta} \phi(\theta,t,\omega)\frac{d\theta}{2\pi}.
\end{equation}
It can be checked that the normalization condition (\ref{norm}) is satisfied. Similarly, the reset density has the Fourier expansion
\begin{eqnarray}
\rho_0(\theta,\omega)=\frac{g(\omega)}{2\pi}\left (1+\sum_{m=1}^{\infty}\left [\rho_{0,m}(\omega)\e^{im\theta}+\mbox{c. c.}\right ]\right ),\nonumber \\
\label{FT0}
\end{eqnarray}
and the corresponding reset order parameter is
\begin{equation}
\label{op2}
Z_{0,1}=R_0\e^{i\psi_0}:=
\int_{0}^{2\pi}d\theta \, \int_{-\infty}^{\infty}d\omega\ \e^{i\theta}\rho_0(\theta,\omega).
\end{equation}
Solving the initial value problem for $\phi$ is then equivalent to solving an infinite hierarchy of equations for the coefficients $\phi_m$:
\begin{align}
&\frac{\partial \phi_m}{\partial t}+im \omega \phi_m +\frac{\lambda m}{2}\left [\phi_{m+1}Z_1 -\phi_{m-1}Z_1^* \right ]+Dm^2\phi_m=-r\phi_m+r \rho_{0,m},
 \label{phin}
\end{align}
with
\begin{equation}
Z_1(t)=\int_{-\infty}^{\infty}g(\omega)  \phi^*_1(\omega,t) d\omega.
\label{1moo}
\end{equation}

The nonlinear FP Eq. (\ref{MVKur0}) without resetting ($r=0$) has been analyzed extensively elsewhere \cite{Sakaguchi88,Strogatz91,Crawford94,Crawford99}. A major focus of these studies is the stability of the incoherent state  $\sigma_0(\theta,\omega,t)=1/2\pi$. Transitions from this state to a partially synchronized state are analyzed by linearizing Eq. (\ref{MVKur0}) about the incoherent state for $r=0$ and solving a non-trivial spectral problem. A more recent development is to assume that the system evolves on a low-dimensional manifold of states first introduced by Ott and Antonsen \cite{Ott08}. The OA manifold is defined in terms of the Fourier series (\ref{FT}) by setting $\phi_n(\omega,t)=z(\omega,t)^n$ for all $n\geq 1$
with $|z(\omega,t)|<1$ and $z(\omega,t)$ analytic in the lower half complex plane. Substituting the OA solution into Eq. (\ref{MVKur0}) with $r=0$ implies that
\begin{align}
&\frac{\partial z(\omega,t)}{\partial t}+i\omega z(\omega,t)+\frac{\lambda R(t)}{2}\left [z(\omega,t)^2\e^{i\psi(t)}-\e^{-i\psi(t)}\right ] =0.
\label{OA1}
\end{align}
Numerical simulations indicate that in the noiseless Kuramoto model ($D=0$), the Ott-Antonsen (OA) manifold acts as an attractor of the dynamics for appropriate choices of the frequency distribution $g(\omega)$. In the specific case of the Cauchy or Lorentzian distribution
\begin{equation}
\label{Lor}
g(\omega)=\frac{\Gamma}{\pi}\frac{1}{\omega^2+\Gamma^2},
\end{equation}
 one can use Eq. (\ref{OA1}) to derive a closed equation for the first circular moment $Z_1=\langle \e^{i\theta}\rangle$, since $Z_1(t)=z^*(i\Gamma,t)$, see also Appendix B.
The OA theory no longer holds in the case of noisy oscillators ($D>0$). However, a perturbation analysis can be developed in the weak noise regime using circular cumulants rather then circular moments \cite{Tyulkina18}. 

In the following we consider two limiting cases of the MV equation with local resetting. The first is the non-interacting phase oscillator model ($\lambda=0$) for which an exact stationary solution can be obtained. The second is the so-called ``noiseless'' Kuramoto model ($D=0$), where we investigate the effects of resetting on the OA dynamics.

\subsubsection{Non-interacting case ($\lambda=0$)} Setting $\lambda=0$ in Eq. (\ref{phin}) yields the non-recurrent equation
\begin{align}
&\frac{\partial \phi_m}{\partial t}+(im\omega +Dm^2+r)  \phi_m =r \rho_{0,m},
 \label{phin0}
\end{align}
which has the explicit solution
\begin{align}
\phi_m(\omega,t)&=\rho_{0,m}(\omega)\e^{-(r+Dm^2+i\omega m)t}+\frac{r\rho_{0,m}(\omega)}{r+Dm^2+i\omega m}\left [1-\e^{-(r+Dm^2+i\omega m)t}\right ]. 
\end{align}
It follows that the stationary density has the Fourier series expansion (\ref{FT}) with
\begin{align}
\lim_{t\rightarrow \infty}\phi_m(\omega,t)=\frac{r\rho_{0,m}(\omega)}{r+Dm^2+i\omega m},
\end{align}
and
\begin{equation}
Z_1=\lim_{t\rightarrow \infty} Z_1(t)=\int_{-\infty}^{\infty}g(\omega)\frac{r\rho_{0,1}^*(\omega)}{r+D-i\omega }d\omega.
\end{equation}
Taking $g(\omega)$ to be the Lorentzian (\ref{Lor}) and closing the contour in the upper half complex plane gives
\begin{equation}
Z_1=\frac{rZ_{1,0}}{r+D+\Gamma}.
\label{zed}
\end{equation}
where we have set $Z_{0,1}=\rho_{0,1}^*(i\Gamma)$. As expected, in the absence of resetting and interactions, the system is in the incoherent state due to a combination of the noise ($D>0$) and the dispersion in natural frequencies ($\Gamma >0$). Switching on local resetting introduces a certain level of order, assuming that the initial state is partially synchronized, that is, $|Z_{0,1}|>0$.

\subsubsection{Noiseless case ($D=0$)} 
\begin{figure}[b!]
\centering
\includegraphics[width=10cm]{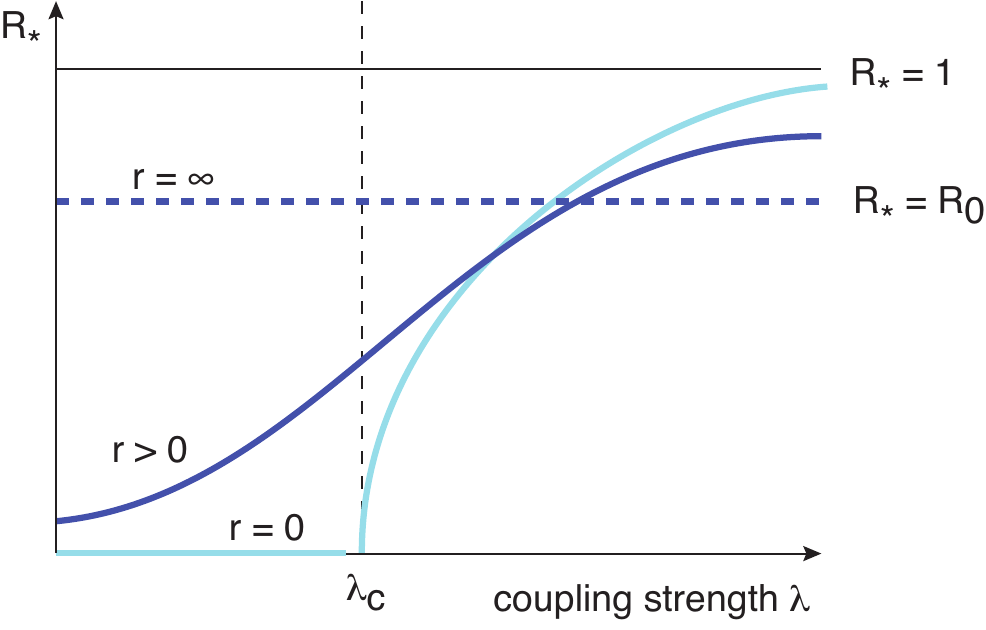} 
\caption{Schematic diagram illustrating the effect of local resetting on the steady state order parameter $R_*=\lim_{t\rightarrow \infty} R(t)$ where $R(t)$ is defined in Eq. (\ref{op}) with $\phi(\theta,t,\omega)$ the solution of the MV Eq. (\ref{MVKur}).}
\label{fig7}
\end{figure}

\begin{figure}[t!]
\centering
\includegraphics[width=10cm]{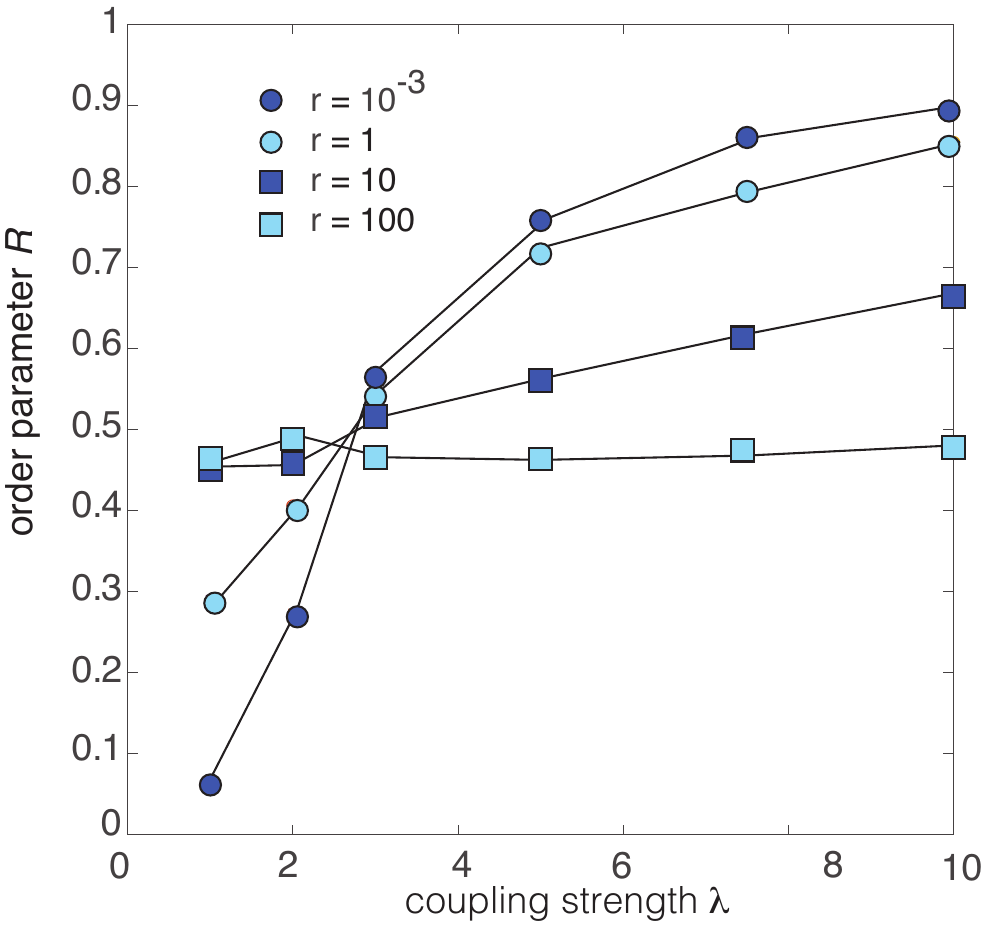} 
\caption{Numerical simulations of $N=400$ Kuramoto oscillators with local resetting. The quenched natural frequencies are generated from the Cauchy distribution (\ref{Lor}) with $\Gamma=1$. Each oscillator evolves according to Eq. (\ref{SDEKur}) and is independently reset to its initial phase at random time intervals generated from an exponential distribution with rate $r$. Data points represent the bare amplitude $R$ of the order parameter $N^{-1}\sum_{j=1}^N\e^{i\theta_j}$ at the end of a single run for different values of $r$ and the coupling strength $\lambda$. The order parameter of the initial density is $R_0=0.496$. We also take $D=0$.}
\label{fig8}
\end{figure}

In Appendix B we use circular cumulants to show how the OA ansatz for $D=0$ breaks down in the presence of local resetting. Unfortunately, it does not appear possible to develop a perturbation analysis in the small $r$ regime along analogous lines to Ref. \onlinecite{Tyulkina18} for $D>0$ and $r=0$. This follows from the observation that all higher-order cumulants are $O(r)$. However, we can gain some insights into the effects of resetting by   assuming that the dynamics remains in a neighborhood of the OA manifold, at least in the small $r$ regime. This yields the approximation
\begin{align}
\frac{dZ_1}{dt}&\approx -\Gamma Z_1+\frac{\lambda}{2}Z_1(1-|Z_1|^2)+r(Z_{1,0}-Z_1),
\end{align}
which reduces to the exact equation for the dynamics on the OA manifold when $r=0$. (Simply set $\omega=i\Gamma$ and $z^*(i\Gamma,t)=Z_1(t)$ in the complex conjugate of Eq. (\ref{OA1}).) Setting $Z_1(t)=R(t)\e^{i\psi(t)}$ and 
decomposing into real and imaginary parts shows that $\psi(t)=\psi_0$ for all $t$ and
\begin{eqnarray}
\frac{d R(t)}{d t}=-\frac{\lambda R(t)}{2}\left [R^2(t)-1 \right ]-(r+\Gamma)R(t)+rR_0.\nonumber \\
\label{RRR}
\end{eqnarray}
The fixed points of Eq. (\ref{RRR}) satisfy the inhomogeneous cubic equation
\begin{equation}
R_*[\lambda R_*^2/2+\Gamma +r-\lambda/2]=rR_0.
\end{equation}

\begin{figure}[t!]
\centering
\includegraphics[width=10cm]{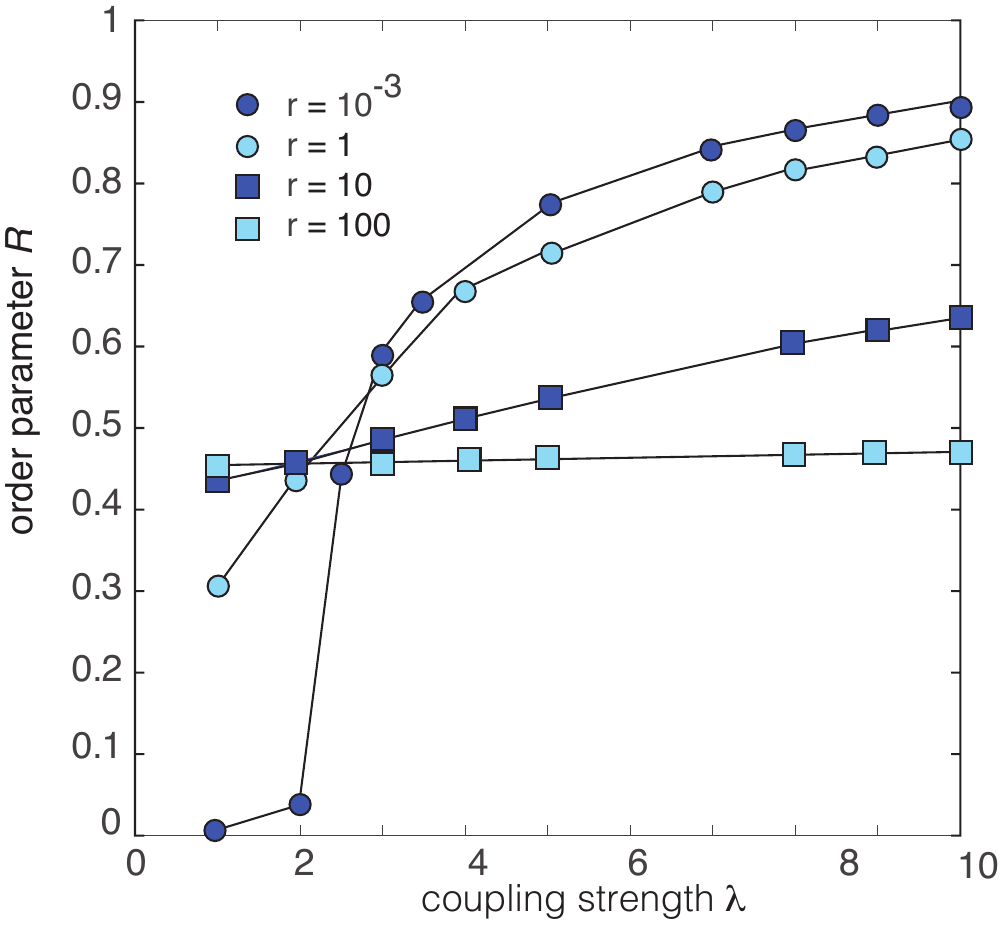} 
\caption{Same as Fig. \ref{fig9} except that $N=4000$ and $R_0=0.457$.}
\label{fig9}
\end{figure}

First consider the classical  case $r=0$. If $\lambda < 2\Gamma$ then there is only a single stable fixed point $R_*=0$ which represents an incoherent state. On the other hand, if $\lambda> 2\Gamma$ then the incoherent state is unstable and there is a stable partially coherent state $R_*=R_{c}(\lambda)\equiv \sqrt{1-2\Gamma/\lambda}$. Hence, the critical coupling $\lambda_c =2\Gamma$ acts as a bifurcation point for the transition from the incoherent state to a partially synchronous state. When $r>0$ there is only one positive root, which is a monotonically increasing function of $\lambda$ for all $\lambda$,
 as illustrated schematically in Fig. \ref{fig7}. This suggests that turning on local resetting results in an ``imperfect bifurcation'' that smooths the transition between the incoherent and partially synchronized states.
Such a solution also  interpolates between the limiting cases $r\rightarrow 0$ and $r\rightarrow \infty$. In the latter case, the last two terms on the right-hand side of Eq. (\ref{MVKur0}) dominate so that $\phi(\theta,t,\omega) \approx \rho_0(\theta,\omega)$ for all $t$, and thus $R_*=R_0$. 

Although we haven't obtained an exact stationary solution for the MV Eq. (\ref{MVKur0}), the above qualitative analysis of the mean field theory captures the behavior of the Kuramoto model quite well even for relatively small $N$. This is illustrated in Figs. \ref{fig8} and \ref{fig9}, where we show the results of numerical simulations for $N=400$ and $N=4000$ oscillators, respectively. We plot the amplitude $R$ for different values of $r$ and $\lambda$. In each case we show the result of a single run, that is, we don't average over multiple realizations of the noise processes. The predicted trends are clearly seen. It can also be checked that fluctuations over many trials are $O(1/\sqrt{N})$, consistent with the law of large numbers.

\subsection{McKean-Vlasov equation: global resetting} In the case of global resetting with $h_j(t)=h(t)=\sum_n\delta(t-T_n)$ for all $j=1,\ldots,N$, we only take expectations of Eq. (\ref{rhoKurc}) with respect to the white noise process. Imposing the mean field ansatz then leads to a stochastic MV equation for
$\Phi(\theta,t,\omega)= \langle  \rho(\theta,t,\omega)  \rangle$: 
\begin{align}
 \frac{\partial \Phi(\theta,t,\omega)}{\partial t} 
&=D\frac{\partial^2}{\partial \theta^2} \Phi(\theta,t,\omega)  -\frac{\partial}{\partial \theta} \bigg (\Phi(\theta,t,\omega) \calV[\theta,t,\omega,\Phi] \bigg )\nonumber \\
 &\quad +h(t)[\rho_0(\theta,\omega)-\Phi(\theta,t,\omega].
\label{MVKurglob}\end{align}
Again we consider the two cases $\lambda=0$ and $D=0$.

\subsubsection{Non-interacting case ($\lambda=0$).} Setting $\lambda=0$ in Eq. (\ref{MVKurglob}) gives the linear FP equation
\begin{align}
 \frac{\partial \Phi(\theta,t,\omega)}{\partial t} 
&=D\frac{\partial^2}{\partial \theta^2} \Phi(\theta,t,\omega) -\omega \frac{\partial}{\partial \theta}  \Phi(\theta,t,\omega)  +h(t)[\rho_0(\theta,\omega)-\Phi(\theta,t,\omega].
\label{FPKurglob}
\end{align}
We could analyze Eq. (\ref{FPKurglob}) along similar lines to the example of overdamped Brownian particles in Sec. IV  by deriving moment equations for the associated probability density functional, and establishing the existence of two-point correlations etc. Here, we consider the Fourier series expansion 
\begin{eqnarray}
\Phi(\theta,t,\omega)=\frac{g(\omega)}{2\pi}\left (1+\sum_{m=1}^{\infty}\left [\Phi_m(\omega,t)\e^{im\theta}+\mbox{c. c.}\right ]\right ),\nonumber \\
\label{FTs}
\end{eqnarray}
which leads to the following set of uncoupled stochastic equations for the Fourier coefficients:
\begin{align}
&\frac{\partial \Phi_m}{\partial t}+(im\omega +Dm^2)\Phi_m=-h(t)  [\Phi_m-\rho_{0,m}].
 \label{phinH}
\end{align}
Set $\Phi_m=A_m\e^{i\varphi_m}$, $\rho_{0,m}=A_{0,m}\e^{i\varphi_{0,m}}$ and equate real and imaginary parts:
\begin{subequations}
\begin{align}
\frac{\partial A_m}{\partial t}+Dm^2A_m&=-h(t)  [A_m-A_{0,m}\cos (\varphi_m-\varphi_{0,m})]\\
\frac{\partial \varphi_m}{\partial t}+\omega m &=-h(t) \frac{A_{0,m}}{A_m}\sin (\varphi_m-\varphi_{0,m})
\end{align}
\end{subequations}
This is equivalent to the piecewise deterministic ODE
\begin{equation}
\frac{dA_m}{dt}=-Dm^2 A_m,\quad \frac{d\varphi_m}{dt}=-\omega_m
\end{equation}
supplemented by the reset rule $A_m(t)\rightarrow A_{0,m}$ and $\varphi_m(t)\rightarrow \varphi_{0,m}$ at the Poisson generated times $t=T_n,\ n\geq 1$. In between reset events, $T_{n}<t<T_{n+1}$, we have
\[A_m(t)=A_{0,m}\e^{-Dm^2(t-T_n)},\quad \varphi_m(t)=\varphi_{m,0}-\omega_m (t-T_n).\]

\begin{figure}[t!]
\centering
\includegraphics[width=10cm]{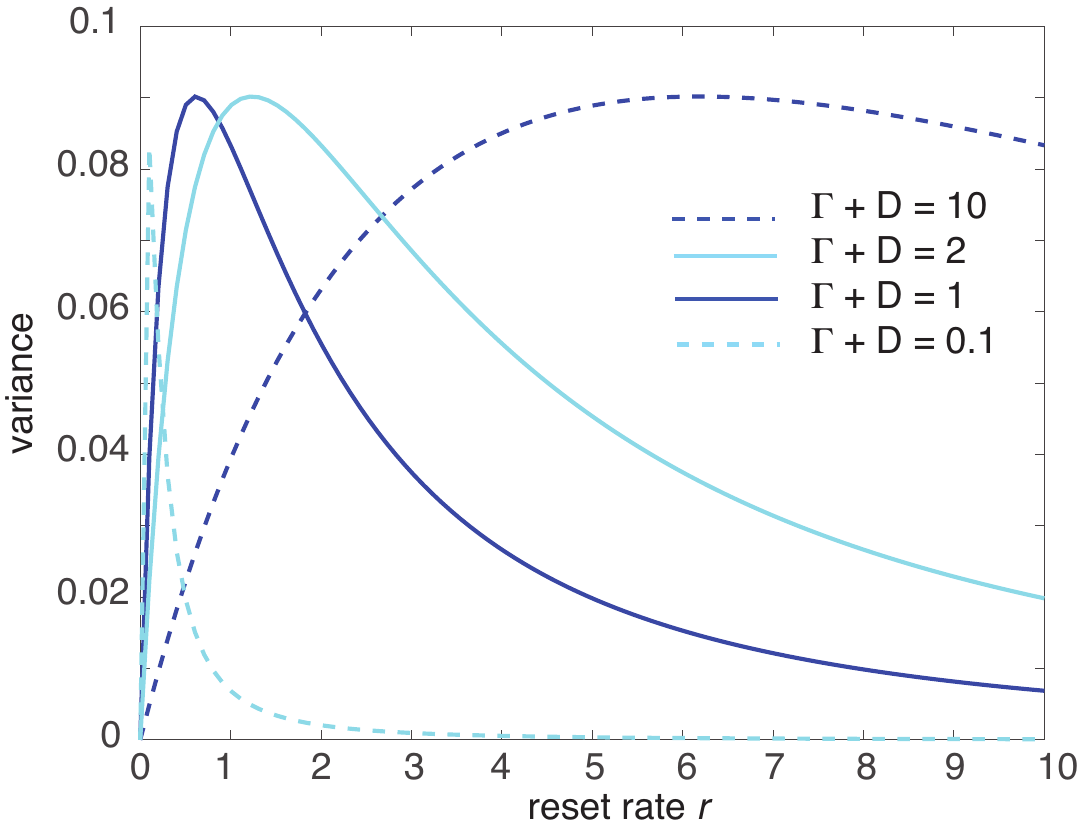} 
\caption{Plot of the variance Var$[\calR]$ in units of $R_0^2$, see Eq. (\ref{var}),  as a function of the reset rate $r$ and various values of $\Gamma+D$. }
\label{fig10}
\end{figure}

In the particular case of the Lorentzian (\ref{Lor}), the first circular moment is $Z_1(t)=\Phi_1^*(i\Gamma,t)$ so that
\begin{align}
\frac{dZ_1}{dt}&=-(\Gamma+D) Z_1+h(t)(Z_{1,0}-Z_1),
\end{align}
Substituting $Z_1=\calR(t)\e^{i\psi(t)}$, equating real and imaginary parts, and taking $\psi(t)=\psi_0$ yields
\begin{equation}
\label{opH}
 \frac{d \calR(t)}{d t}=-(\Gamma+D) \calR(t) +h(t)[R_0-\calR(t)] .
\end{equation}
Introducing the probability density
$p(R,t)dR=\P[\calR(t) \in [R,R+dR]]$,
we have the modified Liouville equation
\begin{equation}
\label{OAH}
\frac{\partial p(R,t)}{\partial t}=\frac{\partial}{\partial R} \bigg ((\Gamma+D)Rp(R,t)\bigg )-rp(R,t) +r\delta(R-R_0),
\end{equation}
The resulting NESS $\lim_{t\rightarrow \infty}p(R,t)=p_0(R)$ is
\begin{equation}
p_0(R)=\frac{r}{\Gamma+D}\frac{1}{R}\left (\frac{R}{R_0}\right )^{r/[\Gamma+D]}\Theta(R_0-R),
\label{p0}
\end{equation}
where $\Theta$ is the Heaviside function. It immediately follows that the steady-state mean of the order parameter $R$ is
\begin{align}
\overline{R}&=\lim_{t\rightarrow \infty}\E[\calR(t)]=\int_0^{\infty}p_0(R)Rdr=\frac{r}{\Gamma+D}\int_0^{R_0} 
\left (\frac{R}{R_0}\right )^{r/[\Gamma+D]}dR=\frac{rR_0}{r+D+\Gamma}.
\end{align}
Comparison with Eq. (\ref{zed}) shows that $\overline{R}$ is identical to the deterministic stationary state $R=|Z_{1}|$ of the non-interacting model with local resetting. In addition, the steady-state variance of the stochastic order parameter $\calR(t)$ is 
\begin{align}
\mbox{Var}[\calR]&=\lim_{t\rightarrow \infty}\E[\calR(t)^2]-\overline{R}^2 =\int_0^{\infty}p_0(R)R^2dR-\overline{R}^2 \nonumber \\
&=\frac{rR_0^2}{r+2(\Gamma+D)} -\left (\frac{rR_0}{r+\Gamma+D} \right )^2\nonumber \\
&=\frac{r(\Gamma+D)^2R_0^2}{(r+\Gamma+D)^2(r+2(\Gamma+D))}.
\label{var}
\end{align}
In Fig. \ref{fig10} we plot $\mbox{Var}[\calR]/R_0^2$ as a function of $r$ for different levels of randomness $\Gamma+D$. We see that the variance is a unimodal function of $r$ with a peak at a value $r_{\rm max}$ that increases with $\Gamma+D$. That is, for higher levels of randomness, the transition from the totally incoherent state $R=0$ to the partially coherent state $R=R_0$ requires a faster rate of resetting.

\subsubsection{Stochastic dynamics on the OA manifold ($D=0$)}

We now set $D=0$ and simplify the stochastic MV equation (\ref{MVKurglob}) by assuming that in  between reset events $(t\neq T_n$) the solution $\Phi(\theta,t)$ lies on the OA manifold
\begin{eqnarray}
\label{OA0}
\Phi(\theta,t,\omega)=\frac{g(\omega)}{2\pi}\left (1+\sum_{m=1}^{\infty}\left [\calZ^m(\omega,t)\e^{im\theta}+\mbox{c. c.}\right ]\right ).\nonumber \\
\end{eqnarray}
This is maintained by resetting provided that the reset density $\rho_0(\theta,\omega)$ also lies on the OA manifold, see Eq. (\ref{FT0}). We thus obtain a stochastic version of Eq. (\ref{OA1}):
\begin{subequations}
\begin{align}
 &\frac{\partial \calZ(\omega,t)}{\partial t}+i\omega \calZ(\omega,t) +\frac{\lambda \calR(t)}{2}\left [\calZ^2(\omega,t)\e^{i\Psi(t)}-\e^{-i\Psi(t)} \right ]=[z_0(\omega)-\calZ(\omega,t)]h(t),
 \label{OA1g}
\end{align}
with
\begin{equation}
\label{opg}
 \calR(t)\e^{i\Psi(t)}:=\int_{0}^{2\pi}d\theta \, \int_{-\infty}^{\infty}d\omega\ \e^{i\theta}\Phi(\theta,t,\omega).
\end{equation}
\end{subequations}
Moreover, taking $g(\omega)$ to be the Lorentzian (\ref{Lor}) such that $\calR(t)\e^{i\Psi(t)}=\calZ^*(-i\Gamma,t)$ and $R_0\e^{i\psi_0}=z_0^*(-i\Gamma)$, we obtain the amplitude
\begin{align}
\frac{d \calR(t)}{d t}&=-\Gamma \calR(t)-\frac{\lambda \calR(t)}{2}\left [\calR^2(t)-1 \right ] +h(t)[R_0-\calR(t)] .
\end{align}
This is a piecewise deterministic equation for $\calR(t)$ that undergoes resetting $\calR(t)\rightarrow R_0$ at the random sequence of times $T_n$ with $\E[T_{n+1}-T_n]=r$. Introducing the probability density
$p(R,t)dR=\P[\calR(t) \in [R,R+dR]]$,
we have the modified Liouville equation
\begin{subequations}
\label{OAreset}
\begin{equation}
\frac{\partial p(R,t)}{\partial t}=-\frac{\partial}{\partial R} \bigg (p(R,t)v(R)\bigg )-rp(R,t) +r\delta(R-R_0),
\end{equation}
with
\begin{equation}
v(R)= -\Gamma R-\frac{\lambda R}{2}\left [R^2 -1 \right ].
\end{equation}
\end{subequations}
We thus recover the master equation introduced in Ref. \onlinecite{Sarkar22} to investigate the effects of global resetting on the stationary state of the Kuramoto model. Here we derived this equation from first principles using the underlying DK Eq. (\ref{rhoKurc}). 

We end by briefly summarizing some of the results obtained for the NESS on the OA manifold\cite{Sarkar22} in order to compare with the case of local resetting.
\medskip

\noindent {\bf Case $\lambda < \lambda_c$.} The time-independent version of Eq. (\ref{OAreset}) has the following explicit solution for $0<R_0\leq 1$ \cite{Sarkar22}:
\begin{align} 
p(R)&=\frac{2r}{\lambda R^3}\left (1+\frac{|\Delta|}{R_0^2}\right )^{\alpha}\left (1+\frac{|\Delta|}{R^2}\right )^{-(1+\alpha)}, \ R\in [0,R_0],\nonumber \\
p(R)&=0,\quad R\in [R_0,1],
\label{pR1}
\end{align}
where
\begin{equation}
\Delta =1-\frac{2\Gamma}{\lambda},\quad \alpha=\frac{r}{2\Gamma -\lambda}.
\end{equation}
Note that there is a discontinuity at $R=R_0$ due to the presence of the Dirac delta function in (\ref{OAreset}). In addition, $\lim_{\lambda\rightarrow 0^+}p(R)=(r/\Gamma) R^{-1}(R/R_0)^{r/\Gamma}=p_0(R)_{D=0}$, where $p_0(R)$ is given by Eq. (\ref{p0}). In a neighborhood of the incoherent state $R=0$ one finds \cite{Sarkar22}
\begin{equation}
p(R)\sim \left (\frac{R}{R_0}\right )^{2\alpha -1},\quad 0<R\ll 1.
\end{equation}
It follows that $p(R)$ diverges as $R\rightarrow 0^+$ when $2\alpha -1<0$, that is, for $r<r_c$ where
\begin{equation}
r_c=\Gamma \left (1-\frac{\lambda}{\lambda_c}\right ),\quad \lambda_c=2\Gamma.
\end{equation}
The normalization condition $\int_0^1p(R)dR=1$ then implies that $p(R)$ is sharply peaked around $R=0^+$. In other words, the system is most likely to be observed in the incoherent state for all resetting rates $r <r_c$. In addition, $p(R)$ is a monotonically decreasing function of $R$. On the other hand, if $r>r_c$ then $p(R)\rightarrow 0$ as $R\rightarrow 0^+$ and is either a unimodal function with a peak at $R=R_m=\sqrt{(2\alpha-1)|\Delta|/3}$ provided that $R_m <R_0$ or a monotonically increasing function in $[0,R_0]$. The main conclusion of this analysis is that resetting can result in a partially synchronized state even though the coupling is below the standard Kuramoto threshold, i. e., $\lambda <\lambda_c$.
\medskip

\begin{figure}[t!]
\centering
\includegraphics[width=10cm]{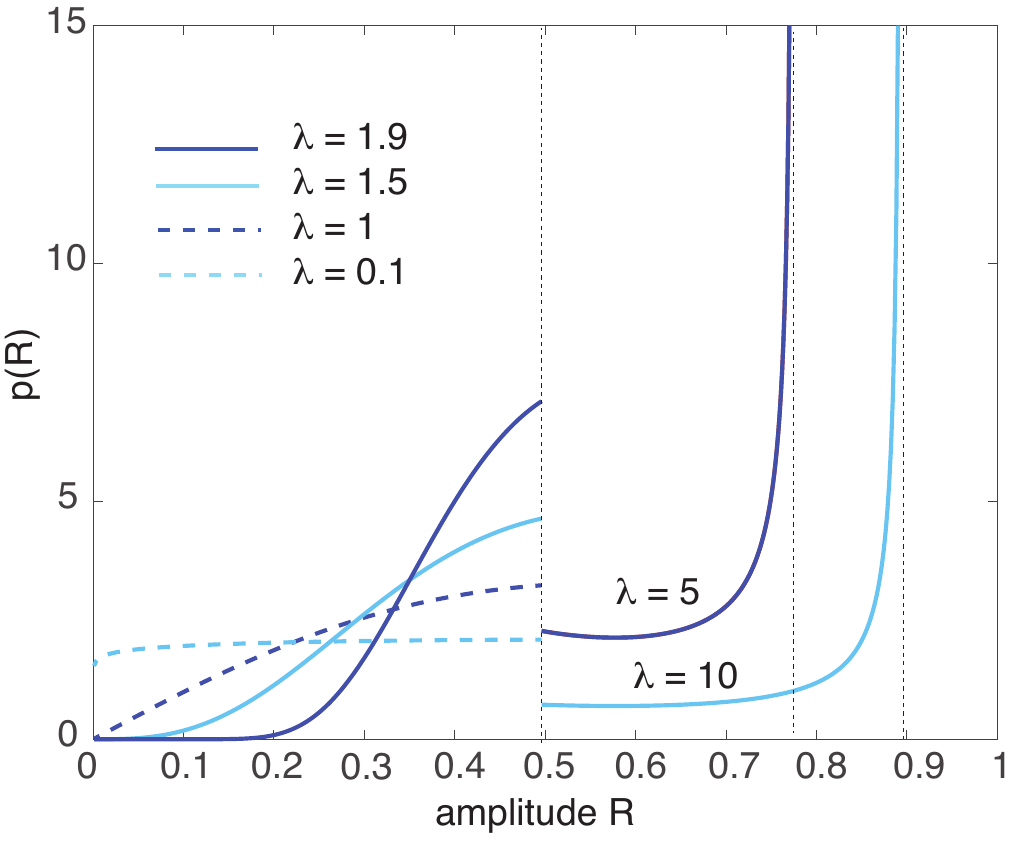} 
\caption{Plot of the NESS $p(R)$ for the amplitude $R$ of the first circular moment, see Eqs. (\ref{pR1})--(\ref{pR3}), for various coupling strengths $\lambda$, with $R_0=0.496$, $D=0$, $\Gamma=1$, and $r=1$.}
\label{fig11}
\end{figure}

\begin{figure}[t!]
\centering
\includegraphics[width=10cm]{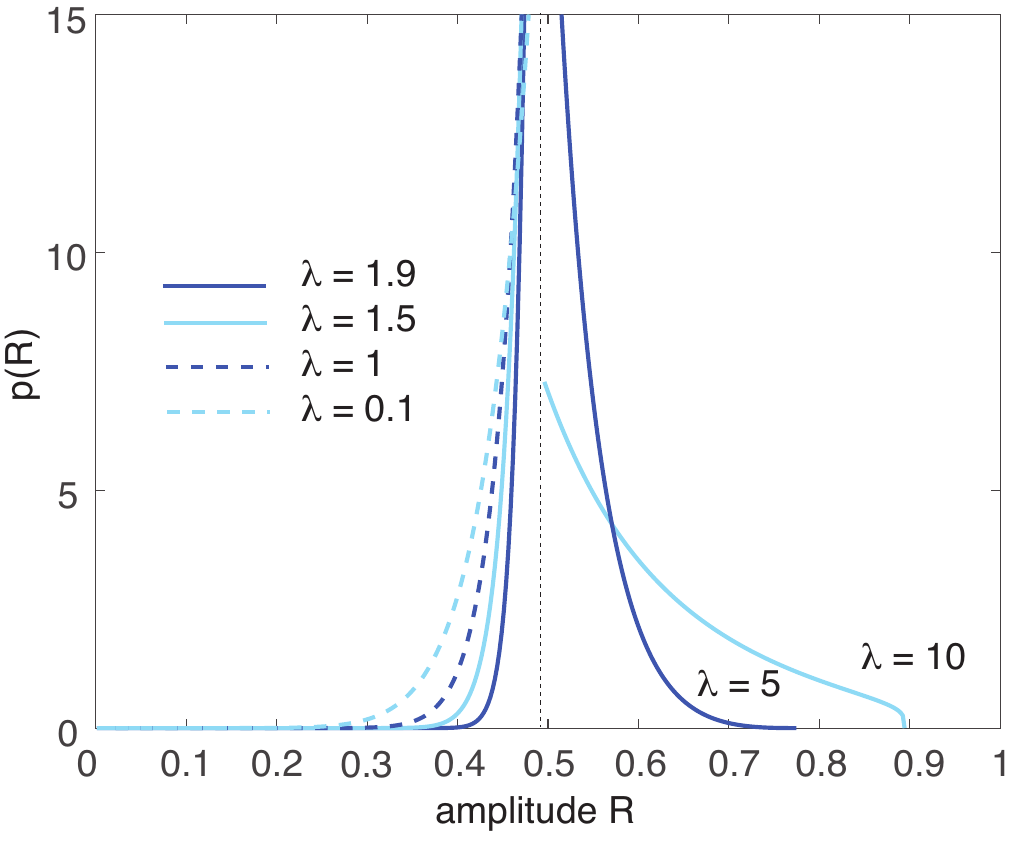} 
\caption{Same as Fig. \ref{fig12} except that $r=0.1$.}
\label{fig12}
\end{figure}

\noindent {\bf Case $\lambda > \lambda_c$.} In the parameter regime that supports a partially synchronized state $R_c(\lambda)=\sqrt{1-2\Gamma/\lambda}$ in the absence of resetting, it can be shown that  if $R_0>R_c(\lambda)$ then
\begin{align} 
p(R)&=\frac{2r}{\lambda R^3}\left (1-\frac{R_c(\lambda)^2}{R_0^2}\right )^{-|\alpha|}\left (1-\frac{R_c(\lambda)^2}{R^2}\right )^{|\alpha|-1} \mbox{ for } R\in [R_c(\lambda),R_0],\nonumber \\
p(R)&=0,\quad R\notin [R_c(\lambda),R_0],
\label{pR2}
\end{align}
whereas
\begin{align} 
p(R)&=\frac{2r}{\lambda R^3}\left (\frac{R_c(\lambda)^2}{R_0^2}-1\right )^{-|\alpha|}\left (\frac{R_c(\lambda)^2}{R^2}-1\right )^{|\alpha|-1}  \mbox{ for } R\in [R_0,R_c(\lambda)],\nonumber \\
p(R)&=0,\quad R\notin [R_0,R_c(\lambda)],
\label{pR3}
\end{align}
when $R_0<R_c(\lambda)$. The qualitative behavior of $p(R)$ differs in the three regimes $r <r_*$, $r=r_*$ and $r=r_*$ with $r_*=\lambda - \lambda_c$, 
However, in all of these cases the most probable state is partially synchronized. See Ref. \onlinecite{Sarkar22} for more details and and a validation of the mean field limit based on numerical simulations. 

In Figs. \ref{fig11} and \ref{fig12} we show example densities $p(R)$ in the case of the initial (reset) amplitude $R_0\approx 0.496$ and the range of parameter values used in Fig. \ref{fig8} for local resetting. In particular, $D=0$, $\Gamma=1$, and $\lambda_c=2$. Fig. \ref{fig11} implies that trial-by-trial variations in $R_*$ are significant for $r=1$ and $\lambda < \lambda_c $ but are localized around $R_c(\lambda)$ when $\lambda>\lambda_c$. On the other hand, Fig. \ref{fig12} shows that if $r=0.1$ then variations are larger when $\lambda > \lambda_c$. Finally, note that the mean order parameter $\overline{R}=\int_0^1 Rp(R)dR$ is a monotonically increasing function of the coupling strength $\lambda$ \cite{Sarkar22}, consistent with the response to local resetting, see Figs. \ref{fig8} and \ref{fig9}.

\setcounter{equation}{0}
\section{Conclusion} In summary, the overarching theme of this paper is that the DK equation for an interacting particle system with local or global stochastic resetting provides a general organizing framework for developing mean field theories. The basic steps are as follows: 
\medskip

\noindent (i) Define an appropriate global density or empirical measure $\rho$. 
\medskip

\noindent (ii) Use It\^o's lemma to construct the DK equation for the global density. This is an exact equation for $\rho$ in the weak sense. 
\medskip

\noindent (iii) Take expectations with respect to the noise processes using a mean field ansatz. This leads to a corresponding MV equation in the thermodynamic limit, which takes the form of a nonlinear nonlocal FP equation. 
\medskip

\noindent We systematically applied this analysis to two canonical examples of an interacting particle system: the Langevin equation for overdamped Brownian particles and the Kuramoto model of coupled phase oscillators. In the former case the global density is $\rho(\x,t)=N^{-1}\sum_{j=1}^N\delta(\x-\X_j(t))$, where $\X_j(t)\in \R^d$ is the position of the $j$th Brownian particle at time $t$, and the corresponding DK equation is (\ref{rhoc}). On the other hand, the global density for the Kuramoto model is $\rho(\theta,t,\omega)=N^{-1}\sum_{j=1}^N\delta(\theta-\Theta_j(t))\delta(\omega-\omega_j)$, where $\Theta_j(t)$ and $\omega_j$ are the phase and natural frequency of the $j$th oscillator, respectively, and the corresponding DK equation is (\ref{rhoKurc}). We also established a major difference between the MV equations for local and global resetting, namely, that the former is deterministic whereas the latter is stochastic. This is a consequence of the fact that global resetting is common to all of the particles and thus induces correlations that cannot be eliminated by taking a mean field limit. 
We obtained an analogous result in our recent work on global density equations for interacting particles that randomly switch between different internal states (local switching) or are subject to a randomly switching environment (global switching) \cite{Bressloff24}. 

We also obtained a number of specific results for stationary solutions of the MV equations.
\medskip

\noindent (a) For Brownian particles with local resetting and Curie-Weiss (quadratic) pairwise interactions, any stationary solution is a nonequilibrium stationary state (NESS) that depends self-consistently on the first moment of the NESS. 
 \medskip
 
 \noindent (b) Local resetting leads to a break down of the Ott-Anderson (OA) theory for the noiseless Kuramoto model ($D=0$). This means that the NESS of the Kuramoto model no longer lies on the two-dimensional OA manifold. Nevertheless, interpolating between the limiting cases $r=0$ and $r=\infty$ indicates that local resetting smooths the phase transition from an incoherent to a partially coherent state. This is also observed numerically.
\medskip
 
 \noindent (c)  Applying OA theory to the stochastic MV equation for the Kuramoto model with global resetting leads to a piecewise deterministic dynamical system with reset on the low-dimensional OA manifold. Stationary solutions are now determined by the NESS $p(R)$ of the stochastic amplitude $\calR(t)$, which has recently been analyzed in Ref. \onlinecite{Sarkar22}.
 \medskip

Mathematically speaking, there are two distinct issues that warrant further study. First, rigorously proving the validity of the mean field ansatz in the presence of local or global resetting by extending previous work in stochastic analysis \cite{Oelsch84,Dai96,Jabin17,Pavliotis21,Chaintron22a,Chaintron22b}. Suppose that in the absence of resetting, the empirical measure of an interacting particle system converges in distribution to the solution of an MV equation written in the general functional form
\begin{equation}
\frac{\partial \phi}{\partial t}={\bm \nabla}\cdot \bigg ( \phi {\bm \nabla}\frac{\delta {\mathcal F}}{\delta \phi}\bigg),
\end{equation}
where ${\mathcal F}[\phi]$ is some free energy functional. Our analysis suggests that the corresponding MV equations with local and global resetting, respectively, are
\begin{equation}
\frac{\partial \phi}{\partial t}={\bm \nabla}\cdot \bigg ( \phi {\bm \nabla}\frac{\delta {\mathcal F}}{\delta \phi}\bigg) +r [\phi_0 -\phi],
\end{equation}
and
\begin{equation}
\frac{\partial \phi}{\partial t}={\bm \nabla}\cdot \bigg ( \phi {\bm \nabla}\frac{\delta {\mathcal F}}{\delta \phi}\bigg) +h(t)[\phi_0-\phi].
\end{equation}
Here $h(t)$ is a Poisson shot noise process and $\phi_0$ is the reset density. 
Second, using the underlying DK equations to develop effective numerical and computational methods for studying large but finite systems with stochastic resetting, along analogous lines to Refs. \onlinecite{Dirr16,Konarovskyi19,Konarovskyi20,Djurdjevac22a,Djurdjevac22b,Fehrman23,Cornalba23}. 

In terms of the specific applications, one possible future direction would be to extend the analysis of interacting Brownian particles with Curie-Weiss interactions and local resetting (Sec. III) to the case of multi-well potentials and their associated phase transitions. One difficulty compared to systems without resetting \cite{Desai78,Dawson83,Pavliotis19} is that we would have to calculate the NESS for the corresponding noninteracting system. There are also a number of possible extensions of the classical Kuramoto model of coupled phase oscillators with resetting. One example is the Kuramoto model with coupling through an external medium, which arises in studies of quorum sensing \cite{Metha12}. Another is the so-called theta model of excitable neurons \cite{Erm86,Bick20}.

\setcounter{equation}{0}
\renewcommand{\theequation}{A.\arabic{equation}}
\section*{Appendix A: Confluent hypergeometric functions}

The confluent hypergeometric equation is
\begin{equation}
\label{cgf1}
x\frac{d^2y(x)}{dx^2}+(c-x)\frac{dy(x)}{dx}-ay(x)=0.
\end{equation}
One solution of this second-order equation is the confluent hypergeometric (or Kummer-M) function 
\begin{align}
 y(x)&={}_1F_1(a,c;x)=M(a,c;x)=1+\frac{a}{c}x+\frac{a(a+1)}{c(c+1)}\frac{x^2}{2!}+\ldots
 \label{M}
\end{align}
for $c\neq 0,-1,-2,\ldots$
The series representation is convergent for all finite $x$. A second linearly independent solution is 
\begin{equation}
y(x)=x^{1-c}M(a+1-c,2-c;x),\quad c\neq 2,3,4,\ldots
\end{equation}
In the limit $|x|\rightarrow \infty$, we have the asymptotic expansion
\begin{equation}
\label{infM}
M(a,c;x)\sim \frac{\Gamma(c)}{\Gamma(a)}\frac{e^x}{x^{c-a}}\left [1+\frac{(1-a)(c-a)}{x}+O(1/x^2)\right ] .
\end{equation}
Hence, $M(a,c;x)$ diverges in the limit $x\rightarrow \infty$ for all $a,c$.
Often the second solution for $x> 0$ is taken to be a linear combination known as the Kummer-$U$ function:
\begin{align}
 U(a,c;x)&=\frac{\pi}{\sin \pi c}\bigg [\frac{M(a,c;x)}{(a-c)!(c-1)!} -\frac{x^{1-c}M(a+1-c,2-c;x)}{(a-1)!(1-c)!}\bigg ].
 \label{U}
\end{align}
This has the asymptotic limit
\begin{equation} 
U(a,c;x)\sim \frac{1}{x^a}\left [1-\frac{a(1+a-c)}{x}+O(1/x^2) \right ]
\end{equation}
as $  x\rightarrow 0$. One can construct an analogous function for $x<0$ given by
\begin{align} 
 \overline{U}(a,c;x) &=\frac{\pi}{\sin \pi c}\bigg [\frac{M(a,c;x)}{(a-c)!(c-1)!} +\frac{x^{1-c}M(a+1-c,2-c;x)}{(a-1)!(1-c)!}\bigg ].\end{align}

The confluent hypergeometric Eq. (\ref{cgf1}) is clearly not self-adjoint with respect to the standard $L^2$ norm. Therefore, it is convenient to define the Whittaker-$M$ function
\begin{equation}
M_{k\alpha}(x)=\e^{-x/2}x^{\alpha+1/2} M(\alpha-k+1/2,2\alpha+1;x),
\end{equation}
which satisfies the self-adjoint equation
\begin{equation}
\frac{d^2M_{k\alpha}(x)}{dx^2}+\left (-\frac{1}{4}+\frac{k}{x}+\frac{1/4-\alpha^2}{x^2} \right )M_{k\alpha}(x).
\end{equation}
The corresponding second solution is the Whittaker-$U$ function
\begin{equation}
W_{k\alpha}(x)=e^{-x/2}x^{\alpha+1/2} U(\alpha-k+1/2,2\alpha+1;x).
\end{equation}
Finally, note that an alternative version of the confluent hypergeometric equation is obtained by performing the change of variables $x=z^2 $ so that $d/dx=(2z)^{-1}d/dz$. Eq. (\ref{cgf1}) then becomes
\begin{equation}
\label{cgf2}
\frac{d^2y(z)}{dz^2}+\left [\frac{2c-1}{z} -2z\right ]\frac{dy(z)}{dz}-4ay(z)=0.
\end{equation}
If we now set $c=1/2$ and perform a second change of variables $z^2=\beta \mu w^2/2$ we obtain Eq. (\ref{h}) under the mapping $w\rightarrow x$.

\setcounter{equation}{0}
\renewcommand{\theequation}{B.\arabic{equation}}
\section*{Appendix B: Break down of the OA ansatz due to local resetting}

Following Ref. \onlinecite{Tyulkina18}, consider the moment generating function
\begin{equation}
F(k,t)=\langle \langle \e^{k\e^{i\theta}}\rangle \rangle =\sum_{m=0}^{\infty} Z_m(t)\frac{k^m}{m!},
\end{equation}
where 
\begin{equation}
Z_m(t)=\left .\frac{\partial^m F(k,t)}{\partial k^m}\right |_{k=0}=\langle \langle \e^{im\theta}\rangle\rangle,
\end{equation}
and
\begin{equation}
\langle \langle  \e^{im\theta}\rangle\rangle=\int_{-\infty}^{\infty}g(\omega) \left [\int_0^{2\pi}\e^{im\theta} \phi(\theta,t,\omega)\frac{d\theta}{2\pi}\right ] d\omega
\label{moo}
\end{equation}
is the $m$th circular moment averaged with respect to the natural frequencies. Further simplification occurs in the case of the Cauchy or Lorentzian distribution (\ref{Lor}),
since contour integration can be used to evaluate the right-hand side of Eq. (\ref{moo}):
\begin{align}
Z_m(t)&=\frac{1}{2\pi i}\oint_C\phi^*_m(\omega,t)\left [\frac{1}{\omega-i\Gamma}-\frac{1}{\omega+i\Gamma}\right ] d\omega =\phi_m^*(i\Gamma,t).
\end{align}
The contour $C$ is obtained by closing the real axis in the upper-half complex plane. As special cases we have
\begin{subequations}
\label{zbar}
\begin{align}
R(t)\e^{i\psi(t)}&=Z_1(t)=\phi_1^*(i\Gamma,t),\\
R_0\e^{i\psi_0}&=Z_{0,1}\equiv \rho_{0,1}^*(i\Gamma).
\end{align}
\end{subequations}
Taking the complex conjugate Eq. (\ref{phin}) and setting $\omega=i\Gamma$, it can be shown that
\begin{align}
&\frac{\partial F}{\partial t}+\Gamma k\frac{\partial F}{\partial k}+\frac{\lambda}{2}\left [k\frac{\partial^2 F}{\partial k^2}Z^*_1(t)-kFZ_1(t)\right ] +Dk\frac{\partial}{\partial k}\left (k\frac{\partial F}{\partial k}\right )=-rF+rF_0,
\label{F}
\end{align}
where
\begin{equation}
F_0(k) =\sum_{m=0}^{\infty} \rho_{0,m}\frac{k^m}{m!}.
\end{equation}

The next step is to introduce the circular cumulants $\chi_m$ via the corresponding cumulant-generating function \cite{Tyulkina18}:
\begin{align}
S(k,t)&=k\frac{\partial \ln F(k,t)}{\partial k}=\frac{k}{F(k,t)}\frac{\partial  F(k,t)}{\partial k}:=\sum_{m=1}^{\infty} \chi_m(t)k^m.
\label{S0}
\end{align}
That is,
\begin{equation}
\chi_m(t)=\frac{1}{m!}\left .\frac{\partial^m S(k,t)}{\partial k^m}\right |_{k=0}.
\end{equation}
For example,
\begin{equation}
\chi_1=Z_1,\ \chi_2=Z_2-Z_1^2,\ \chi_3 = \frac{1}{2}\bigg (Z_3-3Z_2Z-1+2Z_1^3\bigg ).
\end{equation}
Similarly,  the cumulant expansion of the reset density is
\begin{equation}
\chi_{0,m}=\frac{1}{m!}\left .\frac{\partial^m S_0(k)}{\partial k^m}\right |_{k=0},\quad S_0(k)=k\frac{\partial \ln F_0(k)}{\partial k}),
\end{equation}
with $\chi_{0,1}=Z_{0,1}$, see eq. (\ref{op2}).
Differentiating Eq. (\ref{S0}) with respect to $t$ gives
\begin{align}
\frac{\partial S}{\partial t}=\frac{k}{F}\frac{\partial}{\partial k}\left [\frac{\partial F}{\partial t}\right ]-\frac{S}{F}\frac{\partial F}{\partial t}.
\end{align}
Substituting for $\partial F/\partial t$ using (\ref{F}) gives
\begin{align}
&\frac{\partial S}{\partial t}+\Gamma k\frac{\partial S}{\partial k}+\frac{\lambda Z_1^*(t)}{2}k\frac{\partial }{\partial k}\left [k\frac{\partial }{\partial k}\left (\frac{S}{k}\right )+\frac{S^2}{k} \right ]-\frac{\lambda k Z_1(t)}{2}  \nonumber \\
&\quad +Dk\frac{\partial}{\partial k}\left (k\frac{\partial S}{\partial k}+S^2\right )=r[S_0-S]\frac{F_0}{F}.
\label{S}
\end{align}
We have set $S_0=k\partial \ln F_0/\partial k$. Finally, differentiating Eq. (\ref{S}) with respect to $k$ and setting $k=0$ yields an infinite system of equations for the circular cumulants:
\begin{align}
&\frac{d \chi_n}{dt}+n\Gamma \chi_n-\frac{\lambda  Z_1(t)}{2}  \delta_{n,1} + \frac{\lambda  Z_1^*(t)}{2} \left (n^2 \chi_{n+1}+n\sum_{m=1}^n \chi_m\chi_{n+1-m}\right )\nonumber \\
&\quad +D\left (n^2\chi_n+n\sum_{m=1}^{n-1}\chi_{n-m}\chi_m\right )=\left. \frac{r}{n!}\frac{\partial^n }{\partial k^n}\left ([S_0(k)-S(k,t)]\frac{F_0(k)}{F(k,t)}\right )\right |_{k=0}.
\label{hier}
\end{align}

In the case of the classical noiseless Kuramoto model ($D=r=0$), the higher-order cumulant equations decouple from $\chi_1$ so that the the OA manifold defined by the conditions $\chi_n=0$ for all $n\geq 2$ is an invariant manifold of the dynamics. On this manifold  $\phi_n(\omega,t) =z(\omega,t)^n$ with $|z(\omega,t)|<1$ in the Fourier series (\ref{FT}), and \cite{Ott08,Tyulkina18}
\begin{align}
\frac{dZ_1}{dt}&=-\Gamma Z_1+\frac{\lambda}{2}Z_1(1-|Z_1|^2).
\end{align}
However, both the $D$-dependent and $r$-dependent terms in Eq. (\ref{hier}) couple $Z_1$ to higher-order cumulants, indicating a break down of the OA ansatz. In order to focus on the effects of local resetting we set$D=0$ and assume that the reset density lies on the OA manifold, that is, $\chi_{0,m}=0$ for all $m\geq 2$. The analysis is still complicated because local resetting couples the first and higher circular cumulants due to the term on the right-hand side of Eq. (\ref{hier}). Moreover, it is not possible to develop a perturbation expansion in the small-$r$ regime since all of the higher-order cumulants are $O(r)$. Therefore, we proceed by assuming that the dynamics remains in a neighborhood of the OA manifold and drop the contributions from the cumulants $\chi_n$, $n\geq 2$. It follows that the right-hand side of Eq. (\ref{hier}) becomes
\begin{align}
[S_0(k)-S(k)]\frac{F_0(k)}{F(k)}=k[Z_{0,1}-Z_1]\e^{k[Z_{0,1}-Z_1]}.
\end{align}
and the equation for the first circular moment is
\begin{align}
\frac{dZ_1}{dt}&=-\Gamma Z_1+\frac{\lambda}{2}Z_1(1-|Z_1|^2)+r(Z_{1,0}-Z_1).
\end{align}

 \section*{Author declarations}
 \subsection*{Conflict of Interest}
 The author has no conflicts to declare
 \subsection*{Author Contributions}
\begin{small}
\noindent {\bf Paul C. Bressloff}: Conceptualization; Formal analysis; Writing - original draft; Writing - review and editing.
\end{small}
 
 \section*{Data availability}
 
 No data was involved in this study.

\end{document}